\documentclass[tradiabstract]{aa} 

\usepackage{graphicx}
\usepackage{txfonts}

\usepackage{epsf}
\usepackage{rotating}
\usepackage{graphics}

\def \sophie{{\it SOPHIE}}

\def \MS{M$_{\odot}$}
\def \kms{km\,s$^{-1}$}
\def \ms{m\,s$^{-1}$}
\def \1s{$1\,\sigma$}

\def \t0{T$_0$}

\def \cible{HD\,80606}
\def \cibleb{{\cible}b}

\newcommand{\spitzer}{\emph{Spitzer}}

\begin{document}

   \title{Observation of the full 12-hour-long transit of the exoplanet \cibleb\thanks{Based 
            on observations collected with the {\it SOPHIE}  spectrograph on the 1.93-m telescope at
            Observatoire de Haute-Provence (CNRS), France, 
            and with the \textit{Spitzer Space Telescope}, which is operated by the Jet Propulsion Laboratory, 
            California Institute of Technology under a contract with NASA.}}
            
        \subtitle{Warm-\textit{Spitzer} photometry and \textit{SOPHIE} spectroscopy}

   \author{G.~H\'ebrard\inst{1}, 
   	        J.-M.~D\'esert\inst{2,1},
                 R.~F.~D\'{\i}az\inst{1},
                 I.~Boisse\inst{1},
	        F.~Bouchy\inst{1,3}, 
                 A.~Lecavelier des Etangs\inst{1},
                 C.~Moutou\inst{4}, 
                 D.~Ehrenreich\inst{5},
                 L.~Arnold\inst{3},
                 X.~Bonfils\inst{5,7},
                 X.~Delfosse\inst{5},
                 M.~Desort\inst{5},
                 A.~Eggenberger\inst{5},  
                 T.~Forveille\inst{5}, 
                 J.~Gregorio\inst{6},
                 A.-M.~Lagrange\inst{5},
                 C.~Lovis\inst{7},
                 F.~Pepe\inst{7},
                 C.~Perrier\inst{5}, 
                 F.~Pont\inst{8}, 
                 D.~Queloz\inst{7},
	       A.~Santerne\inst{4},
                 N.~C.~Santos\inst{9},
                 D.~S\'egransan\inst{7},
                 D.~K.~Sing\inst{8},
	        S.~Udry\inst{7}, 
                 A.~Vidal-Madjar\inst{1}
}


   \institute{
Institut d'Astrophysique de Paris, UMR7095 CNRS, Universit\'e Pierre \& Marie Curie, 
98bis boulevard Arago, 75014 Paris, France 
\email{hebrard@iap.fr}
\and
Harvard-Smithsonian Center for Astrophysics, 60 Garden Street, Cambridge, MA 02138, USA
\and
Observatoire de Haute-Provence, CNRS/OAMP, 04870 Saint-Michel-l'Observatoire, France
\and
Laboratoire\,d'Astrophysique\,de\,Marseille,\,Univ.\,de\,Provence,\,CNRS\,(UMR6110),\,38\,r.\,F.\,Joliot\,Curie,\,13388\,Marseille\,cedex\,13,\,France
\and
Laboratoire d'Astrophysique de Grenoble, Universit\'e Joseph-Fourier, CNRS (UMR5571), BP53, 38041 Grenoble cedex 9, France
\and
CROW-observatory Portalegre, Portugal, and Atalaia.org group, Portugal
\and
Observatoire de Gen\`eve,  Universit\'e de Gen\`eve, 51 Chemin des Maillettes, 1290 Sauverny, Switzerland
\and
School of Physics, University of Exeter, Exeter, EX4 4QL, UK 
\and
Centro de Astrof{\'\i}sica, Universidade do Porto, Rua das Estrelas, 4150-762 Porto, Portugal
}

   \date{Received TBC; accepted TBC}
      
  \abstract{We present new observations of a transit of the 111.4-day-period exoplanet \cibleb.
Due to this long orbital period and to the orientation of the eccentric orbit ($e=0.9$), the  
\cibleb's transits last for about 12 hours. This makes practically impossible the observation 
of a full transit from a given ground-based observatory. 
Using the \spitzer\ \textit{Space Telescope} and its IRAC camera on the post-cryogenic mission, 
we performed a 19-hour-long photometric observation of 
\cible\ that covers the full transit of 13-14 January 2010 as well as off-transit references 
immediately before and after the event. We complement this photometric data by new 
spectroscopic observations that we simultaneously performed with \sophie\ at Haute-Provence 
Observatory. This provides radial velocity measurements of the first half of the transit that was 
previously uncovered with spectroscopy. 
This new data set allows the parameters of this singular planetary system to be significantly 
refined. We obtained a planet-to-star radius 
ratio $R_\mathrm{p}/R_* = 0.1001 \pm 0.0006$ that is more accurate but 
slightly lower than the one measured from previous ground observations in the optical. 
We found no astrophysical interpretations able to explain such a difference 
between optical and infrared radii; 
we rather favor 
underestimated systematic uncertainties, maybe in the ground-based composite light curve.
We detected a feature in the \spitzer\  light curve that could be due to a stellar spot.
We also found a transit timing about 20~minutes 
earlier than the ephemeris prediction; this could be caused by actual transit timing variations 
due to an additional body in the system, or again by underestimated systematic uncertainties. 
The actual angle between the spin-axis of \cible\ and the normal to the planetary orbital plane 
is found to be near $40^{\circ}$ thanks to the fit of the Rossiter-McLaughlin anomaly, 
with a sky-projected value $\lambda = 42^{\circ} \pm 8^{\circ}$. This allows scenarios with 
aligned spin-orbit to be definitively rejected. 
Over the twenty planetary systems with measured spin-orbit angles, a few of them are 
misaligned; this is probably the signature of two different evolution scenarios for 
misaligned and aligned systems, depending if they experienced or not gravitational 
interaction with a third body. 
As in the case of \cible, most of the planetary systems 
including a massive planet are tilted; 
this could be the signature 
of a separate evolution scenario for massive planets in comparison with 
Jupiter-mass~planets.}


  \authorrunning{H\'ebrard et al.}
\titlerunning{Observation of the full 12-hour-long transit of  \cibleb}

   \maketitle


\section{Introduction}

\begin{figure}[h] 
\begin{center}
\includegraphics[scale=0.417]{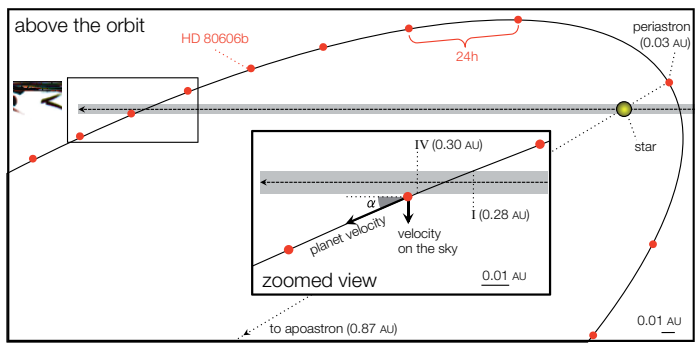}
\caption{Schematic view of the \cible\ system. The red dots show the positions 
of the planet each 24 hours.}
\label{fig_david}
\end{center}
\end{figure}

Among the  more than 400 extrasolar planets that have been found so
far, the giant planet orbiting the  
G5 star \object{\cible} is certainly 
a unique case. Its eccentricity is particularly high: 
$e=0.93$. Only one known planet possibly has a higher eccentricity, 
namely HD\,20782b; however its high $e$-value is still to be confirmed 
as it stands on one measurement only (O'Toole et al.~\cite{otoole09}).
The comet-like orbit of \cibleb\ was well established in its discovery 
paper by Naef et al.~(\cite{naef01}) and has been largely confirmed by 
subsequent observations.
Together with its 111.4-day period, the high eccentricity of \cibleb\ 
put it on an 
extreme orbit: during its revolution, the planet experiences the strongly 
irradiated regime of a "hot Jupiter" at periastron (0.03~AU), and milder 
conditions at apoastron (0.87~AU), around which it spends most of its 
time, and where the planet approaches the inner boundary of the 
habitable zone (see Fig.~\ref{fig_david}).

As long as the inclination $i$ of the orbit with the line of sight was unknown, 
those parameters originally implied a probability of 1\% for the planet to 
be transiting. 
In spite of this tenuous probability, an amazing chance makes the planet 
\cibleb\ actually transits its parent star, as seen from Earth, each 111.4~days.
This is particularly advantageous as numerous crucial studies can be 
performed using photometry or spectroscopy when a planet passes in 
front its parent star (planetary transits) or behind it (planetary eclipses), 
especially in this case where the host star is bright ($V=9.1$) and 
nearby ($d=60$~pc).
The fortunate transiting nature of \cibleb\ was established in February~2009 from the 
detection of a transit reported from ground observations, independently by Moutou 
et al.~(\cite{moutou09}) from photometric and spectroscopic data, and by 
Garcia-Melendo \& McCullough~(\cite{garcia09}) and Fossey et al.~(\cite{fossey09})
from photometric measurements. Additional observations of transits were later reported by 
Winn et al.~(\cite{winn09a}) and Hidas et al.~(\cite{hidas10}).
The February~2009 observations followed the planetary eclipse discovery reported 
a few months earlier by Laughlin et al.~(\cite{laughlin09}) from \spitzer\ photometric 
observations at 8\,$\mu$m during a 30-hour interval around the periastron. 
Among the  known transiting planets, \cibleb\ has the longest period and the most eccentric 
orbit. The second most extreme transiting planet is HD\,17156b ($P=21.2$~days and 
$e=0.67$). Furthermore, \cibleb\ is also the most distant planet from its parent stars 
when it transits: 0.3~AU, against 0.05~AU or less for all other known cases\footnote{See 
also the 95-day-period transiting exoplanet CoRoT-9b announced after the submission 
of the present paper (Deeg et al.~\cite{deeg10}).}.

In addition to the photometric detection of the \cibleb's transit, Moutou et al.~(\cite{moutou09})
presented its spectroscopic detection using the Rossiter-McLaughlin 
effect, from 
radial velocities measured with the \sophie\ spectrograph at Haute-Provence Observatory (OHP), 
France. The Rossiter-McLaughlin effect is an apparent distortion of the stellar lines profile due to the transit 
of the planet in front of the rotating star. From the \sophie\ measurements, Moutou et 
al.~(\cite{moutou09}) have shown the first evidence for a spin-orbit misalignment, i.e. the orbital 
plane of the planet \cibleb\ is not perpendicular to the spin-axis of its host star.
Using additional photometric data of the February 2009 event allowing a better constraint 
on the transit duration together with a combined analysis of the whole data set, 
Pont et al.~(\cite{pont09}) refined the parameters of the system. They confirmed the spin-orbit 
misalignment from the Rossiter-McLaughlin distortion detected with \sophie\ 
and provided a measurement of the sky-projected angle between 
the planetary orbital axis and the stellar rotation axis: $\lambda\sim50^{\circ}$, with the 
confidence interval [$14^{\circ}-111^{\circ}$] -- see 
also Gillon~(\cite{gillon09a}). Thanks to new photometric and spectroscopic 
observations of the June 2009 transit, Winn et al.~(\cite{winn09a}) thereafter 
reduced even more the confidence interval to [$32^{\circ}-87^{\circ}$]. Thus, 
the spin-orbit misalignment of the \cible\ system is now well established.
\cible\ is the component of a wide binary with HD\,80607; the projected 
separation of the system is about 1000~AU. The peculiar orbit of 
\cibleb\ could thus result from Kozai mechanism and tidal dissipation 
(see, e.g., Wu~\& Murray~\cite{wu03}), which can pump
the eccentricity and the inclination.

Due to the long orbital period and to the orientation of the eccentric orbit, the 
duration of the transit of \cibleb\ is about 12-hour long. 
This should be compared to the transit duration 
of other known transiting exoplanets, which typically lasts less than five hours. 
The transit duration of \cibleb\ is even longer than 
transits of Mercury or Venus through the Solar disk as seen from the Earth.
It is thus practically impossible that a full transit of \cibleb\ matches the duration 
of an observation night from ground. In addition, data secured before and after the transit 
are mandatory to obtain an accurate transit light curve, so the full sequence for \cible\ 
lasts longer than a night for ground observations. 
Observing an entire transit of this exoplanet is thereby challenging and
only portions of a transit could be observed from a 
given ground-based telescope. This was the case of all the observation 
campaigns reported above which covered only fractions of transits.
These fragmented observations induce significant uncertainties in the 
parameters derived from their~fit.

We present here the first full photometric observation of a transit of 
\cibleb. We secured it on 13-14 January 2010 with the \spitzer\ space observatory 
using the IRAC camera at 4.5\,$\mu$m in post-cryogenic mission.
Thanks to its Earth-trailing heliocentric orbit (Scoupe et al.~\cite{spot06}), 
\spitzer\ allowed us to continuously observe during 19~hours, enabling the coverage 
of the whole 12-hour-long transit, as well as off-transit references immediately 
before and after the event. We complement this photometric data by new spectroscopic
observations that we simultaneously performed with \sophie\ at OHP. 
This provides radial velocity measurements of the first half of the transit, a part  
that was up to now uncovered  with spectroscopy. Indeed, 
observing the full 12-hour-long transit is even more difficult in spectroscopy
than in photometry, as the amplitude of the Rossiter-McLaughlin for \cible\ 
is about $10$\,\ms\ whereas Northern instruments able to measure radial velocities with 
the required accuracy are sparse. 
We also performed a ground-based photometric monitoring of \cible\ during January 2010. 
All together, the data of this observational campaign allows the parameters of this 
planetary system  to be additionally~refined.

The  observations and data reduction are presented in Sect.~\ref{sect_spitzer}
and~\ref{section_sophie} for \spitzer\ and \sophie\  respectively.
The ground-based photometry in presented in Sect.~\ref{sect_ground_photom}.
The analyses and the results are presented in Sect.~\ref{sect_analysis}, before discussion and 
conclusion in Sect.~\ref{sect_discussion} and ~\ref{sect_conclusion}.

\section{\textit{Spitzer} photometry}
\label{sect_spitzer}

\subsection{Observations}

We obtained \spitzer\ Director's Discretionary Time (DDT program \#540) to observe the 
January 2010 transit of \cibleb. This transit was the first observable with \spitzer\ since the 
discovery of the transiting nature of \cibleb\ in February 2009. As \spitzer\ has exhausted its 
cryogen of liquid coolant on 15 May 2009, our observations were performed during the first 
months of the \spitzer's warm mission. Only the two first infrared channels of the IRAC camera 
(Fazio et al.~\cite{fazio04})
are available in the post-cryogenic \spitzer. They are centered at 3.6 and 4.5\,$\mu$m and 
cannot be observed simultaneously. We chose to observe only in one of the two 
channels in order to avoid repointing the telescope during the transit. 
This reduces overheads and 
ensures that the target remains on the same part of the
detector during the full observation sequence. We opted for channel~2 at 4.5~$\mu$m since 
it has the lowest noise properties. This wavelength also has a smaller limb-darkening effect
for the star. 
We did not use pointing dithering,~here again to maintain the target as much as possible on 
the same location on the detector in order to reduce systematic effects due to imperfect 
flat-field corrections and intra-pixel sensitivity~variations.

We used the Subarray mode of IRAC which is well adapted for 
bright targets. Only a $32\times32$-pixel part of the detector is used in this mode; 
this covers a small $38\times38\;\mathrm{arcsec}^2$ field of view (pixel size of 1.2~arcsec),
compared to the IRAC Stellar mode that uses the full $256\times256$-pixel 
field. As the stellar companion HD\,80607 is located only 17\,arcsec to the East of \cible, 
putting the two targets on this small field of view would have imply that their point spread 
functions (PSF) fall near the edges of the detector. Such configuration is risky for accurate 
photometry. We preferred to let HD\,80607 off the field of view in order to maintain \cible\ 
on the detector, far from its edges. So we chose to put our target at the default 
pointing position in the center of the Subarray field of view. This position is 
not on nor right next to any known hot pixels.

The observations were secured between 2010 January 13 at 18\,h and January 14 at
13\,h (UT). We acquired 2488 consecutive images during a total of 19 hours. 
Each image was split up into 64 frames of 0.36~second exposure each, taken 
back-to-back. 
We obtained a total of 159232 frames 
during 15.9-hours effective integration time with an overhead of 2-second 
between each images (15\,\% overheads in total).
Such high efficiency is reached thanks to the use of the Subarray mode and could not 
be achieved in Stellar mode for bright targets requiring short exposure times.
With a flux density of $\sim200$~mJy at 4.5\,$\mu$m for \cible, frame 
exposure time of 0.36~sec clearly avoids saturation of the pixels. The intensities
recorded in the brightest central pixel are around 10\,000~electrons.  

\subsection{Data reduction}
\label{sect_spitzer_data_reduction}

We used the BCD files (Basic Calibrated Data) of each of the 159232 frames as 
they are produced by the \spitzer/IRAC pipeline. It includes corrections for dark 
current, flat fielding, pixels non-linearity, and conversion to flux~units.
To extract the light curve, we used tools and methods we developed in 
D\'esert et al.~(\cite{desert09}, \cite{desert10}). 
We find the center of the PSF of the star to a precision of 
$0.01$\,pixel using the DAOPHOT-type Photometry Procedures,
\texttt{GCNTRD}, from the IDL Astronomy 
Library\footnote{{\tt http://idlastro.gsfc.nasa.gov/homepage.html}}, 
which computes the
stellar centroid by Gaussian fitting. We used the \texttt{APER} routine 
to extract the raw flux of \cible\ on each frame from the 
computation of a simple aperture photometry using a radius of $4.0$\,pixels, 
which optimize the signal-to-noise ratio of the transit light curve. 
The flux integrated on these 50~pixels is  $\sim52\,500$~electrons.
It has been corrected from the background level of $14.40\pm0.05$~electrons/pixel
that has been estimated from a sky annulus with radii of 9 to 12 pixels.
The centroid of HD\,80607 is located outside the field of view but a small contribution
of its flux is detected on an edge of the detector. When estimating the background level,
we took care to use only pixels where the HD\,80607's flux is negligible. It is also negligible
by comparison to the background level in the 4-pixel radius we used for the \cible\ photometry.
Finally, the uncertainty in the background level is negligible for all the results presented
in this paper.

The \spitzer/IRAC photometry is known to be systematically affected by
the so-called \textit{pixel-phase effect}. This effect produces an oscillation of the
measured raw light curve due to the \spitzer\ telescope jitter and the intra-pixel
sensitivity variations on the IRAC detector (see, e.g., Charbonneau et al.~\cite{charbonneau05},
Reach et al.~\cite{reach06}, Morales-Calder\'on et al.~\cite{morales06}, Ehrenreich
et al.~\cite{ehrenreich07}, D\'esert et al.~\cite{desert09},~\cite{desert10}).
Measurements of the centroid position of the target on the detector and its variations
could be used to de-correlate the pixel-phase effect on the light curve.
We use here the method presented in D\'esert et al.~(\cite{desert09}); 
it has the form 
$F_{\mathrm{corr}}=F[1+K_1(x-x_0)+K_2(x-x_0)^2+K_3(y-y_0)+K_4(y-y_0)^2+K_5(x-x_0)(y-y_0)]$,
where $F$ and $F_{\mathrm{corr}}$ are the fluxes of the star before and after the pixel-phase 
effect correction, and $(x-x_0)$ and $(y-y_0)$ are the position in pixel of the source centroid on 
the detector with respect to the  pixel pointing position, located on $[x_0,y_0]$.
Our determination of the centroid position shows a $\pm0.05$-pixel oscillation with a period
of $\sim70$~minutes and a linear drift during the 19-hour sequence of $0.1$ and  
$0.2$~pixel in the $x$- and $y$-direction,~respectively.

\begin{figure}[h] 
\begin{center}
\vspace{0.2cm}
\includegraphics[scale=0.55]{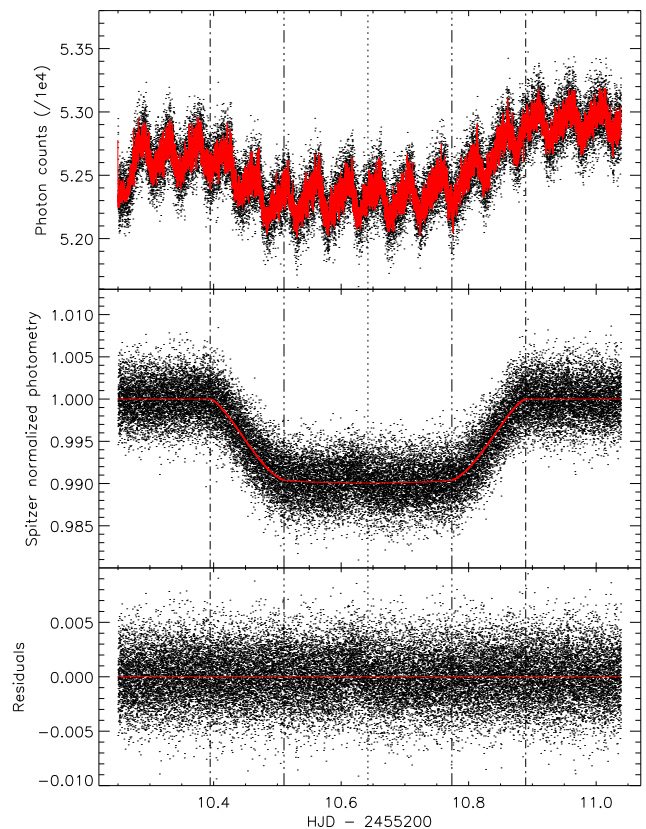}
\caption{\spitzer\ photometry of \cible. From top to bottom, the three panels show 
respectively the raw photometry, the photometry after correction of the pixel-phase 
and ramp effects, and the residuals. The~black points are the 
measurements (binned per 5) and the red solid lines are the~fits. 
The vertical dotted lines show the mid-transit, the vertical dot-dashed 
lines show the first and fourth contacts, and the vertical dot-dot-dashed lines show the second 
and third contacts. }
\label{fig_spi}
\end{center}
\end{figure}

In addition to the 64 frames of the image \#\,873 that are corrupted, we iteratively
selected and trimmed 1037~outliers by comparison to the fit of the light curve with 
a transit model. Frames were considered as outliers when they were above 
10 to 3\,$\sigma$, this value being reduced by 0.1-$\sigma$ steps at each iterations. 
We binned the obtained \emph{transit light curve} by a factor of five in
order to obtain a better computing efficiency without losing
information for the pixel phase effect. 
Most of the results presented
below were obtained using the binned transit light curve.

We tested several radii for the aperture photometry, several areas for the background measurement, 
several procedures for the centroid determination and the outliers rejections.
The adopted procedure reported above is the one producing the smallest errors, but all of them
produced similar~results.

The upper panel of Fig.~\ref{fig_spi} shows the raw \spitzer\ light curve of \cible\ after
this extraction. It clearly shows flux variations with a period of $\sim70$~minutes that are
due to the pixel-phase effect and telescope jitter at this period. 
Its peak-to-peak amplitude represents $\sim1$\,\%\ of the 
flux, which is larger than the effect seen in the channel~2 of IRAC during 
the cryogenic \spitzer. 
Additional variations at higher frequency and with a lower 
amplitude than the one of the 70-minute oscillation are seen as well, 
as we fit each individual frame. They could 
be due to short-term jitter of the satellite which apparently are 
not periodic. Filtering these high-frequency variations does not 
significantly change our results.
Finally, a slope in the out-of-transit baseline is also seen. 
Such detector ramp was observed in channels~3 and 4 in pre-cryogenic \spitzer\ 
but was not as significant in channel~2 at that time
(see, e.g., Deming et al.~\cite{deming05}, Knutson et al.~\cite{knutson07}, 
D\'esert et al.~\cite{desert09},~\cite{desert10}).
We discarded the first thousands of frames  that were the most affected 
by the ramp effect (we tested several limits; see Sect.~\ref{sect_best_parameters}) 
then normalized the light-curve using a time-dependent function 
with the form $F_{baseline} = A_0 + A_1 \times t$ where $F_{baseline}$ 
is the target flux out of transit and $t$ is the time. 
We also tried higher-degree polynomials and logarithmic baselines; 
this did not improve significantly the fit.
The linear correction we adopted is not perfect; remaining uncertainties in the 
actual shape of the baseline introduces errors in the system parameters
derived from the fit (see Sect.~\ref{sect_best_parameters}). 
The baseline and pixel-phase-effect parameters are fitted simultaneously 
with the transit-related parameters (see Sec.~\ref{sect_free_parameters}); 
this allows these effects and their uncertainties to be taken into account 
in the transit parameters determination. The middle
panel of Fig.~\ref{fig_spi} shows the light curve after removal of these  instrumental effects and
the lower panel shows the residuals to the fit.

\section{\textit{SOPHIE} radial velocities}
\label{section_sophie}

\subsection{Observations}

The first half of the January 2010 transit of \cibleb\ was visible from Europe so we 
managed to observe it with the \sophie\ spectrograph at the 1.93-m telescope of 
Haute-Provence Observatory in South of France.
\sophie\ is a cross-dispersed, environmentally stabilized echelle spectrograph dedicated 
to high-precision radial velocity measurements (Perruchot et al.~\cite{perruchot08}, 
Bouchy et al.~\cite{bouchy09}). 
Since the discovery of the transit in February 2009, this was the first time 
that this part of the transit was observable from an observatory with high-precision 
spectroscopic capabilities. Unfortunately, the 1.93-m telescope started a technical break
for maintenance and upgrades 
in November~2009 that was extended up to February~2010. Due to the importance of 
this event, some observations could nevertheless be performed thanks to the support of 
the OHP staff, wich was mandatory due to the ongoing works on the telescope. 

The observations were secured as part of the second sub-program of the \sophie\ Consortium
(Bouchy et al.~\cite{bouchy09}, H\'ebrard et al.~\cite{hebrard10b}). The night of the transit, 
13 January 2010, observations could start just before 23h (UT) after technical issues were solved 
and clouds disappeared, and had to be stopped 4.5~hours later due to cloudy weather. 
A 5.5-hour sequence of good \sophie\ reference observations of \cible\ could also be 
performed the 15 January 2010 night. 24 and 33 exposures were secured during the two 
nights respectively. The exposure times ranged between 5 and 20 minutes, with typical 
values around 9 minutes; we tuned it in order to maintain a constant signal-to-noise 
ratio per pixel of $\sim58$ at 550~nm despite the weather changes (seeing and absorption).

The measurements were performed with the same setup as the one we used for our observation 
of the February 2009 transit (Moutou et al.~\cite{moutou09}).
We used the fast-read-out-time mode of the CCD detector in order to minimize overheads.
Observations were secured in \textit{high-resolution} mode allowing the resolving power 
$\lambda/\Delta\lambda=75000$ to be reached. The first optical-fiber aperture was
put on the target and the second one on the sky; it allows us to check that no diffuse light 
was polluting the \cible\ spectra in these Moonless nights. Wavelength calibrations with 
a thorium lamp were performed with a $\sim2$-hour frequency each night, allowing the 
interpolation of the spectral drift of~\sophie\ for the time of each exposure. A few exposures 
were performed with simultaneous thorium-lamp light in the second aperture~to allow for 
simultaneous wavelength calibration; those extra-calibration did not improve significantly 
the radial velocity~accuracy.

\begin{figure}[h] 
\begin{center}
\includegraphics[scale=0.55]{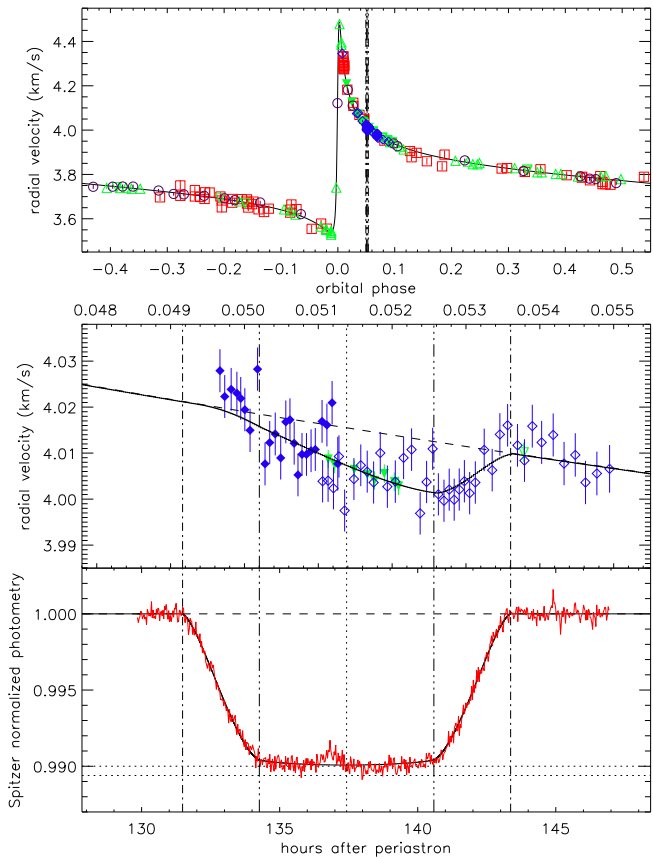}
\caption{Radial velocities and photometry of \cible\ as function of the orbital phase or 
the time after periastron.
\textit{Upper panel:} radial velocities as measured with 
\textit{ELODIE} (red open squares, Moutou et al.~\cite{moutou09}),
\textit{HRS} (purple open circles, Wittenmyer et al.~\cite{wittenmyer09}),
\textit{HIRES} (Winn et al.~\cite{winn09a}) 
pre- and post-upgrade (green open upward and downward triangles, 
respectively) and around the June-2009 transit (green filled downward triangles), and 
\textit{SOPHIE} during the February-2009 and January-2010 transits (blue open and filled 
diamonds, respectively).
\textit{Middle panel:} same as above, but enlarged around the transit phase. 
\textit{lower panel:} \spitzer\ photometry during the January-2010 transit. The 
\spitzer\ data are binned 
per 250, i.e. one point each 100~sec.
The two horizontal dotted-lines show the absorption depth expected with the 
value $R_\mathrm{p}/R_* = 0.100$ that we get and the value $R_\mathrm{p}/R_* = 0.103$ 
measured in the visible by Pont et al.~(\cite{pont09}) and Winn et al.~(\cite{winn09a}).
\textit{On the three panels}, the dashed lines show the models without transit and the solid 
lines show the models with transit (Rossiter-McLaughlin anomaly for radial velocities, 
and absorption feature for the light curve). The parameters of the fits are reported in 
Table~\ref{table_parameters}.
The vertical dotted lines show the mid-transit, the vertical dot-dashed 
lines show the first and fourth contacts, and the vertical dot-dot-dashed lines show the second 
and third contacts. The uncertainty on the timing of the mid-transit is 1.5~minute (corresponding 
to $9\times 10^{-6}$ in orbital phase), and about two times larger for the timing of the four contacts.}
\label{fig_transit_all}
\end{center}
\end{figure}

\subsection{Data reduction}
\label{sect_sophie_reduction}

We used the \sophie\ pipeline (Bouchy et al.~\cite{bouchy09}) to 
extract the spectra from the detector images, 
to cross-correlate them with a G2-type numerical mask, 
then to fit the cross-correlation functions (CCFs) by Gaussians to get the radial velocities 
(Baranne et al.~\cite{baranne96}, Pepe et al.~\cite{pepe02}).
Each spectrum produces 
a clear CCF, with a  $7.82 \pm 0.03$~km\,s$^{-1}$ full width at half 
maximum and a contrast representing $48.8 \pm 0.4$~\%\ of the continuum.
The accuracy of the measured radial velocities is 
typically around 4\,\ms. This includes photon noise 
(typically $\sim2$~\ms), wavelength calibration ($\sim2$~\ms), and 
guiding errors ($\sim3$~\ms) that produce motions of the input image within the 
fiber (Boisse et al.~\cite{boisse09}).  We also re-reduced the \sophie\ observations 
presented in Moutou et al.~(\cite{moutou09}) in order to have a uniform data set 
reduced with the same version of the pipeline. There was no significant change by 
comparison with the data presented in Moutou et al.~(\cite{moutou09}), except for 
one of the reference exposures performed out of the transit for which the correction 
due to the Moon pollution was significantly improved. \sophie\ measurements of 
\cible\ performed in February 2009 and January 2010 have the same properties, 
except the better signal-to-noise ratios for shorter exposure times obtained in 
2010, which is largely due to the primary mirror of the telescope that was realuminized 
in October 2009.

The \sophie\ radial velocities of \cible\ are plotted in Figs.~\ref{fig_transit_all} 
and~\ref{fig_RM} together with other data~sets.

In the \sophie\ spectra, the cores of the large \ion{Ca}{ii} H \& K
absorption lines of \cible\ at  3934.8\,\AA\ and 3969.6\,\AA\ 
show no chromospheric emissions. The level of the \ion{Ca}{ii} emission 
corresponds to $\log{R'_\mathrm{HK}}=-5.3\pm 0.1$ according to the \sophie\ 
calibration (Boisse et al.~in preparation). 
For a G-type star ($B-V=0.76$) with this level of activity, Santos et al.~(\cite{santos00}) 
predict a dispersion below 5~\ms\ for the activity-induced stellar jitter. 
According to Noyes et al.~(\cite{noyes84}) and Mamajek \& Hillenbrand~(\cite{mamajek08}), 
this level of activity implies a stellar rotation period $P_\mathrm{rot} > 50$~days.

\section{Ground-based photometry}
\label{sect_ground_photom}

In order to study its intrinsic variability, \cible\ was also observed several nights before and 
after the January~2010 transit. Observations during the night of the transit were prevented by bad 
weather conditions in the observing site, located in Portalegre, Portugal. We gathered a total of 
nearly 23~hours of observations in 10~nights spread from December 3, 2009 to January 25, 
2010. The equipment used is a 12-inch $f/5.5$  MEADE LX200 telescope, and a 
SBIG ST8XME 9-micron-pixel CCD camera. The field is $28 \times 18.7\;\mathrm{arcmin}^2$ 
and the pixel scale is $\sim 1.1$~arcsec/pixel. Observations were taken through~a~Bessell 
I filter, using integration times between 40 and 90~seconds.

The frames were reduced using standard IRAF routines and aperture photometry was 
obtained for \cible, its companion HD\,80607 and two additional reference stars located 
in the same field, located about 7 and 11.5~arcmin away from \cible. The size of the photometric 
aperture was varied on each night, depending on seeing conditions, in order to obtain the 
highest possible signal-to-noise ratio.

In every night, the flux ratio between \cible\ and HD\,80607 exhibits a smaller dispersion 
than the ratio of \cible\ with any of the other two reference stars, or a combination of HD\,80607 
and the additional reference stars. Furthermore, the light curves obtained for HD\,80607 -- using the reference 
stars -- also exhibit a larger dispersion than the flux ratio between \cible\ and HD\,80607. 
Therefore, the limits to the intrinsic variability of \cible\ were set using only HD\,80607 as reference star.

The root mean square 
for the different nights range from about 2.2 to 4.8~mmag, depending mainly on 
the weather conditions. During each night, no effect was seen due to variation of the sky level, 
or the position of the stars on the chip or the airmass. On the other hand, the nightly mean 
of our observations exhibits a dispersion of about 1.7~mmag around its mean value. 

Similar photometric observations of \cible\ were conducted the weeks around the January 2010
transit with the CCD camera at the 120-cm telescope of Haute-Provence Observatory, using the 
setup used by Moutou et al.~(\cite{moutou09}). The filter used was  r$_{Gunn}$ with a neutral density
and we did not defocus. 
Photometric observations started on 6~January and ended on 23~January, with a total of 
eight sequences ranging from 30 to 120~minutes per night; in total, 187~images have been 
acquired and analyzed. Exposure times of 10 to 120~seconds have been used to account 
for varying transparency. 
An observation of the transit
during the January-13 night could be performed with this instrument, on a
coverage similar to this of the \sophie\ observations reported in Sect.~\ref{section_sophie}.
This ground-based photometry of a part of the transit will be studied in a 
forthcoming~paper.

The flux has been extracted from aperture photometry using the \texttt{GCNTRD} and 
\texttt{APER} procedures (Sect.~\ref{sect_spitzer_data_reduction}) on a 8-pixel 
radius ($0.69$~arcsec/pixel). The background 
has been estimated from a sky annulus with radii of 10 to 12 pixels.
The root mean square of flux variations ranges from 1.0 to 3.1~mmag 
per night, and the residual fluctuation has a standard deviation of 3.2~mmag when the transit 
night is excluded. This is significantly larger than what was observed in Portalegre, which 
may indicate a real tendency of rotational modulation, or a significant level of systematics. 
No attempt was done to correct for the long-term behavior in the data, which appears to be 
compatible with the expected rotational period of the star (as discussed in Sect.~\ref{sect_discuss_RM}).

We conclude therefore that \cible\ is photometrically stable at the level of a few mmag in 
the optical range, in the timescale of several weeks.

\section{Analysis}
\label{sect_analysis}

We fitted this whole data set in order to refine the system parameters. As a full transit was observed
with \spitzer, possible systematic effects due to the combination of transit portions secured 
with different ground-based 
instruments are expected to be reduced here. In addition, together with the new 
radial velocities secured at phases previously uncovered, the constraints on the 
spin-orbit angle would be better.

\subsection{Combined fit}
\label{sect_combined_fit}

\subsubsection{Method}

We first performed a combined fit of our \spitzer\ photometry
of the January-2010 transit together with the available radial velocities of \cible.
We also included in this combined fit the timing constraint 
on the eclipse as measured from previous \spitzer\ measurements by Laughlin et 
al.~(\cite{laughlin09}) and re-analyzed by Gillon~(\cite{gillon09a}).
As those data are not accurate enough to allow the ingress and egress of the eclipse 
to be significantly measured, we used the estimated epochs of the mid-time of these 
two events: HJD$\,=2\,454\,424.700 \pm 0.005$ and $2\,454\,424.775 \pm 0.005$.

We did not include in this combined 
fit the radial velocity measurements secured during and near the transits. 
Indeed, the Rossiter-McLaughlin observations do constrain the projected 
stellar rotational velocity $V \sin I_*$ and the sky-projected angle $\lambda$ 
between the planetary orbital axis and the stellar rotation axis, but they do 
not constrain significantly the parameters that we measure here in the 
combined fit. The analysis of the Rossiter-McLaughlin data is presented below 
in Sect.~\ref{sect_RM} and takes into account the results of the combined 
fit presented here in Sect.~\ref{sect_combined_fit}.
Thus, the combined fit uses the radial velocities secured with the instruments 
\textit{ELODIE} at OHP (Moutou et al.~\cite{moutou09}), 
\textit{HRS} at HET (Wittenmyer et al.~\cite{wittenmyer09}), 
and \textit{HIRES} at Keck 
(Winn et al.~\cite{winn09a}). This covers a 9.5-year span. 
The \sophie\ data are used only for the Rossiter-McLaughlin fit.
We note that the \textit{ELODIE} and \textit{SOPHIE} radial velocities are absolute 
heliocentric whereas those from \textit{HIRES} and \textit{HRS} are~relative.

\subsubsection{Transit light curve of planets on eccentric orbit}
\label{sect_ecc_LC}

To calculate the transit light curve using a given set of orbital parameters (period, orbital inclination, 
semi major axis in unit of stellar radii, eccentricity and longitude of periastron), we calculated the 
sky projected distance between the planetÕs and the starÕs center in unit of stellar radius; this last result 
is then directly input to the Mandel \& Agol~(\cite{mandel02}) algorithm with limb darkening 
coefficients. 
From a theoretical model (Kurucz~\cite{kurucz79}), the three non-linear limb-darkening coefficients
at 4.5\,$\mu$m as defined by Sing~(\cite{sing10}) has been derived 
with $T_\mathrm{eff} = 5500$\,K and $\log g=4.5$: 
$c_2=0.89502981$,     $c_3= -1.1230710$ and     $c_4=0.46541027$. 
The limb darkening is low at this wavelength; D\'esert et al.~(\cite{desert09})
have shown that the uncertainties in the coefficients  
have no significant effects on the parameters derived from the transit light~curve.

It is worth to note that the speed of the light in the system 
has to be taken into account when comparing the times of the transit, the eclipse and 
the periastron. The transit takes place when the planet is at $\sim0.29$\,AU from the star
(see~Fig.~\ref{fig_david}); this implies an apparent advance of about 2.5~minutes for the transit. 
Similarly, the planet is at $\sim0.03$\,AU from the star at the eclipse, which implies a 
delay of 15~seconds. Also because of the 
planet-observer distance decreases during the transit, the speed of light correction makes the transit 
appears to last 8~seconds less than its really does. In all the procedure, we implement this speed 
of light correction. 
    
\begin{figure}[h] 
\begin{center}
\includegraphics[scale=0.51]{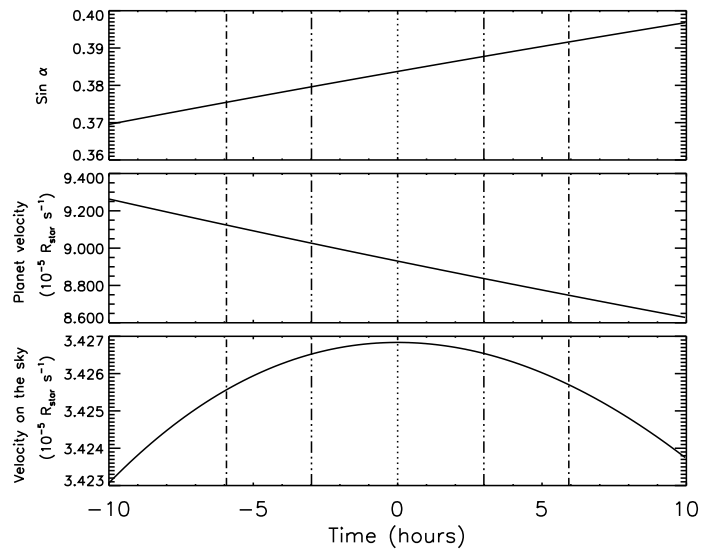}
\caption{Plot of the orbital velocity of the planet and orbit direction during the 
transit. $\alpha$ being the angle between the direction of the planet 
motion and the line of sight, $\sin \alpha$ is the fraction of the planet motion 
projected onto the sky (upper panel). Although the planet velocity significantly 
decreases during the transit (middle panel), the change of the orbital direction 
has the opposite effect such that the planet velocity on the sky is nearly constant 
(bottom panel). Even more, the planet velocity on the sky is nearly symmetrical 
around the center of the transit, reinforcing the apparent symmetry of the transit light curve. 
On the three panels, the vertical dotted lines show the mid-transit, the vertical dot-dashed 
lines show the first and fourth contacts, and the vertical dot-dot-dashed lines show the second 
and third contacts. 
Figure~\ref{fig_david} shows these parameters drawn on a sketch.
}
\label{fig_alain}
\end{center}
\end{figure}    
    
The orbit of \cibleb\ is highly eccentric, and the transit takes place $\sim$5.7~days after the periastron 
when the planet is rapidly moving away from the star. During the transit, the star-planet distance increases 
by about 5.8\,\%\ between the first and the fourth contact, and the orbital velocity decreases by 
about 4.1\,\%\ (Fig.~\ref{fig_alain}). With that in mind, {\it a priori}, this would suggest that the transit light 
curve could be highly asymmetrical with, for instance, an egress lasting longer than the ingress. However, 
during the transit, the direction of the planet motion also varies by about the same amount (upper panel 
of~Fig.~\ref{fig_alain}). If we define $\alpha$ as the angle between the direction of the planet motion and 
the line of sight, $\sin \alpha$ is the fraction of the planet motion projected onto the sky
(see~insert in~Fig.~\ref{fig_david}). Between the first 
contact and the last contact, $\sin \alpha$ increases by about 4.3\,\%, exactly compensating for the 
decrease of the orbital velocity. As a result, the planet velocity on the sky is nearly constant. In addition 
to the low variation of the apparent planet velocity on the sky during the transit, it appears that the time 
variation of this velocity is nearly symmetrical around the center of the transit (bottom panel 
of~Fig.~\ref{fig_alain}). Consequently, the apparent velocity during the egress is extremely close to the 
velocity during the ingress, the later being only 0.004\,\%\  smaller than the former. This reinforces the 
apparent symmetry of the transit light curve. With the parameters of the best fit, the ingress duration 
(time between first and second contact) is predicted to last only half a second more than the egress 
duration (time between third and fourth contact) while both lasts about 2 hours and 45 minutes. 

This surprising symmetry for the transit light curve of a planet on a highly eccentric orbit can be explained 
by considering the projection on the sky of the gravitational force from the star to the planet. Because 
the transited star is the same as the attracting star, during the transit the component of the 
gravitational force projected on the sky is 
 close to zero. As a consequence, the apparent 
velocity of the planet projected on the sky is nearly constant. Even more, for various positions of the 
planet, the projected force is also symmetrical around the star center, explaining the symmetry of the 
time variation of the projected velocity around the center of the transit.
As a conclusion, in contrary to the common-sense idea, in any configuration 
the transit of a planet in front of its parent star is expected to be highly symmetric, 
even for extremely eccentric~orbits (see also Winn~\cite{winn10a}). 
Our \spitzer\ observations confirm this result
in the case of the highly eccentric \cibleb's orbit.

\subsubsection{Free parameters}
\label{sect_free_parameters}

Our combined fit of the radial velocities, the \spitzer\ transit light curve and 
the eclipse timing includes 19 free parameters. We list 
them below, classified in four categories depending on their nature and the 
data set that constrain them:

\begin{enumerate}

  \item 
two free parameters constrained by photometry only:
\begin{itemize}
  \item $R_\mathrm{p} / R_*$, the ratio of the planetary and stellar radii;
  \item $a / R_*$, the semimajor axis in units of stellar radius;
\end{itemize}

  \item 
six free parameters constrained both by radial velocities and~photometry:
\begin{itemize}
  \item $a \cos i /R_*$, the ``standard'' impact parameter, which is different (for an eccentric orbit) of the actual 
  impact parameter $b=d_t \cos i /R_*$, where $d_t$ is the star-planet distance at mid-transit and $i$ is the 
  inclination of the orbit;
  \item $P$, the orbital period of the planet, which is proportional to $(a/R_*)^3$, and to the inverse 
  of the stellar density $\rho_*$ from the Kepler third law;
  \item $T_0$, the epoch of the periastron of the~planet;
  \item $e\cos \omega$ and $e\sin \omega$, that constrain the two correlated parameters 
  $e$ (the eccentricity of the planetary orbit) and $\omega$ (the longitude of its periastron);
  \item $K\sqrt{1-e^2}$, that depends both on the eccentricity $e$ and on the semi-amplitude  $K$ of the 
  radial velocity variations -- $K$ actually is constrained by radial velocities only, and is proportional to 
  $M_\mathrm{p}$ and $(M_*)^{2/3}$, $M_\mathrm{p}$ and $M_*$ being the planetary 
  and stellar masses;
  \end{itemize}

  \item 
four free parameters constrained by radial velocities only:
\begin{itemize}
  \item $V_{n=1\rightarrow4}$, the center-of-mass radial velocities for each of the four radial 
  velocity data set used, namely \textit{ELODIE} (Moutou et al.~\cite{moutou09}), 
  \textit{HRS} (Wittenmyer et al.~\cite{wittenmyer09}), and \textit{HIRES} 
  whose pre- and post-upgrade data are considered as two different data sets (Winn et 
  al.~\cite{winn09a}); 
\end{itemize}

  \item 
seven free parameters linked to the \spitzer\ light curve extraction  (see Sect.~\ref{sect_spitzer_data_reduction}):
\begin{itemize}
  \item the two parameters $A_{j=1\rightarrow2}$  for the baseline;
  \item the five parameters $K_{i=1\rightarrow5}$ for the pixel-phase  effect.
\end{itemize}

\end{enumerate}

The direct, absolute measurements of $M_*$ and $R_*$, and consequently those of 
$M_\mathrm{p}$ and $R_\mathrm{p}$, are not feasible from this fit; none of those 
four parameters is measurable independently of the other ones from our data set. 
One exception could be the stellar mass that could be directly determined 
through the semi-major axis $a$ thanks the third Kepler law. Indeed, by measuring 
the delay of the eclipse mid-time and the advance of the transit's one 
that are due to the time the light takes to propagate through the \cible\ 
system ($\sim2.5$~minutes difference between those two events by 
comparison to the ephemeris, see
Sect.~\ref{sect_ecc_LC}), the semi-major axis could theoretically be directly measured
in Astronomical Units. The uncertainties in the transit and eclipse measured 
mid-times are however of the order of 1.5 and 4~minutes respectively, 
which implies an accuracy on $a$ of the order of 0.5~AU with this approach; 
this is not constraining here. 

Stellar evolution models remain thus mandatory to estimate~$M_*$.  
Moutou et al.~(\cite{moutou09}) used isochrones to get $M_* = 0.98 \pm 0.07$\,M$_\odot$
from $T_{\mathrm{eff}} = 5574\pm50$~K, $\log g = 4.45 \pm0.05$ and $[$Fe/H$]=0.43$\,dex. 
Such accuracy is typical for stellar evolution models, from which it is difficult to predict stellar 
masses at better than $\pm10$~\%\ (Fernandes \&\ Santos~\cite{fernandes04}). 
Using additional constraints from the transit light curve, 
Pont et al.~(\cite{pont09}) and Winn et al.~(\cite{winn09a}) estimated the stellar mass 
of \cible\ to $0.97 \pm 0.04$ and $1.05 \pm 0.032$~\MS, respectively. The 2-$\sigma$
disagreement between the two estimates shows that we probably reach here the limit 
of the accuracy that is now achievable. In the following, we adopt the conservative interval 
$M_* = 1.01 \pm 0.05$~\MS\ that takes into account those two studies.

\begin{table}[h]
  \centering 
  \caption{Parameters for the \cible\ system}
  \label{table_parameters}
\begin{tabular}{lcc}
\hline
\hline
Parameters & Values and 1-$\sigma$ error bars & Unit \\
\hline
\multicolumn{3}{l}{{\hspace{-0.25cm}}free parameters constrained by photometry only:} \\
$R_\mathrm{p}/R_*$			&	$0.1001 \pm 0.0006$			\\
$a/R_*$						&	$97.0 \pm 1.6$			\\
\hline
\multicolumn{3}{l}{{\hspace{-0.25cm}}free parameters constrained both by photometry and radial velocities:} \\
$a \cos i /R_*$					&	$1.238 \pm 0.011$	\\
$P$ 							& $111.4367 \pm 0.0004$				&   days 	\\
$T_0$ (periastron)				& $2\,455\,204.916 \pm 0.004$		&   HJD 	\\
$e \cos \omega$				& $ 0.4774 \pm 0.0018$			\\
$e \sin \omega$				& $-0.8016 \pm 0.0017$			\\
$K \sqrt{1-e^2}$				& $171.1 \pm 0.5$				&   \ms	\\
\hline
\multicolumn{3}{l}{{\hspace{-0.25cm}}free parameters constrained by radial velocities only:} \\
$V_{\textit{ELODIE}}$  						& $3.7888\pm 0.0023$ 			&   \kms   \\
\hspace{0.5cm} $\sigma(O-C)$			& 17.8		&   \ms	\\
$V_{\textit{HRS}}$  							& $-0.0193 \pm 0.0019$ 			&   \kms   \\
\hspace{0.5cm} $\sigma(O-C)$			& 6.3		&   \ms	\\
$V_{\textit{HIRES}\mathrm{,\,pre-ugrade}}$  		& $-0.1845 \pm 0.0010$ 			&   \kms   \\
\hspace{0.5cm} $\sigma(O-C)$			& 5.5		&   \ms	\\
$V_{\textit{HIRES}\mathrm{,\,post-ugrade}}$  		& $-0.1827 \pm 0.0007$ 			&   \kms   \\
\hspace{0.5cm} $\sigma(O-C)$			& 2.4		&   \ms	\\
\hline
\multicolumn{3}{l}{{\hspace{-0.25cm}}directly derived parameters:} \\
$b$									&	$0.808 \pm 0.007$				\\
$i$									&	$89.269 \pm 0.018$		&   $^{\circ}$\\
$T_t$ (transit mid-time)					& $2\,455\,210.6420 \pm 0.0010$		&   HJD 	\\
Transit duration $T_{1-4}$				& 	$11.88 \pm 0.09$	& h \\	
Transit duration $T_{1-2} =T_{3-4}$			& 	$2.78 \pm 0.10$	& h \\	
$e$									& $0.9330 \pm 0.0005$			\\
$\omega$ 							& $300.77 \pm 0.15$				&   $^{\circ}$ \\
$K$									& $475.3 \pm 2.0$				&   \ms	\\
$\rho_*$								&	$1.39 \pm 0.07$				&   g/cm$^3$\\
\hline
$T_e$ (eclipse mid-time)					& $2\,454\,424.736 \pm 0.003$$\ddag$ &   HJD 	\\
Eclipse duration $T_{1-4}$				& 	$1.85 \pm 0.14$$\ddag$	& h \\
\hline
\multicolumn{3}{l}{{\hspace{-0.25cm}}derived parameters assuming a $M_\star$-value:} \\
$M_\star$	&	$1.01 \pm 0.05$$^\dagger$	&   M$_\odot$   \\
$R_\star$								&	$1.007 \pm 0.024$			&   R$_\odot$ \\
$M_\textrm{p}$							&	$4.08 \pm 0.14$	&   M$_\mathrm{Jup}$\\
$R_\textrm{p}$							&	$0.981 \pm 0.023$				&   R$_\mathrm{Jup}$ \\
$\rho_{_\textrm{p}}$						&	$5.4 \pm 0.4$				&   g/cm$^3$\\
$a$									&	$0.455 \pm 0.008 $	& AU \\	
\hline
\multicolumn{3}{l}{{\hspace{-0.25cm}}Rossiter-McLaughlin effect:} \\
$V \sin I_*$ 				& $1.7 \pm 0.3$	&   \kms   \\
$\lambda$				& $42 \pm 8$	&   $^{\circ}$\\
$V_{\textit{SOPHIE}\mathrm{,\,transit\,February\,09}}$  			& $3.9162 \pm 0.0013$ 			&   \kms   \\
\hspace{0.5cm} $\sigma(O-C)$			& 4.3		&   \ms	\\
$V_{\textit{HIRES}\mathrm{,\,post-ugrade,\,transit\,June\,09}}$  	& $-0.1795 \pm 0.0011$ 			&   \kms   \\
\hspace{0.5cm} $\sigma(O-C)$			& 0.7		&   \ms	\\
$V_{\textit{SOPHIE}\mathrm{,\,transit\,January\,10}}$  			& $3.9018 \pm 0.0013$ 			&   \kms   \\
\hspace{0.5cm} $\sigma(O-C)$			& 5.2		&   \ms	\\
\hline
\multicolumn{3}{l}{$\dagger$: combined value from Pont et al.~(\cite{pont09}) and Winn et  al.~(\cite{winn09a})} \\
\multicolumn{3}{l}{$\ddag$: from Laughlin et al.~(\cite{laughlin09})} \\
\end{tabular}
\end{table}

\subsubsection{Best parameters and error bars}
\label{sect_best_parameters}

\begin{figure}[h] 
\begin{center}
\vspace{0.8cm}
\includegraphics[scale=0.52]{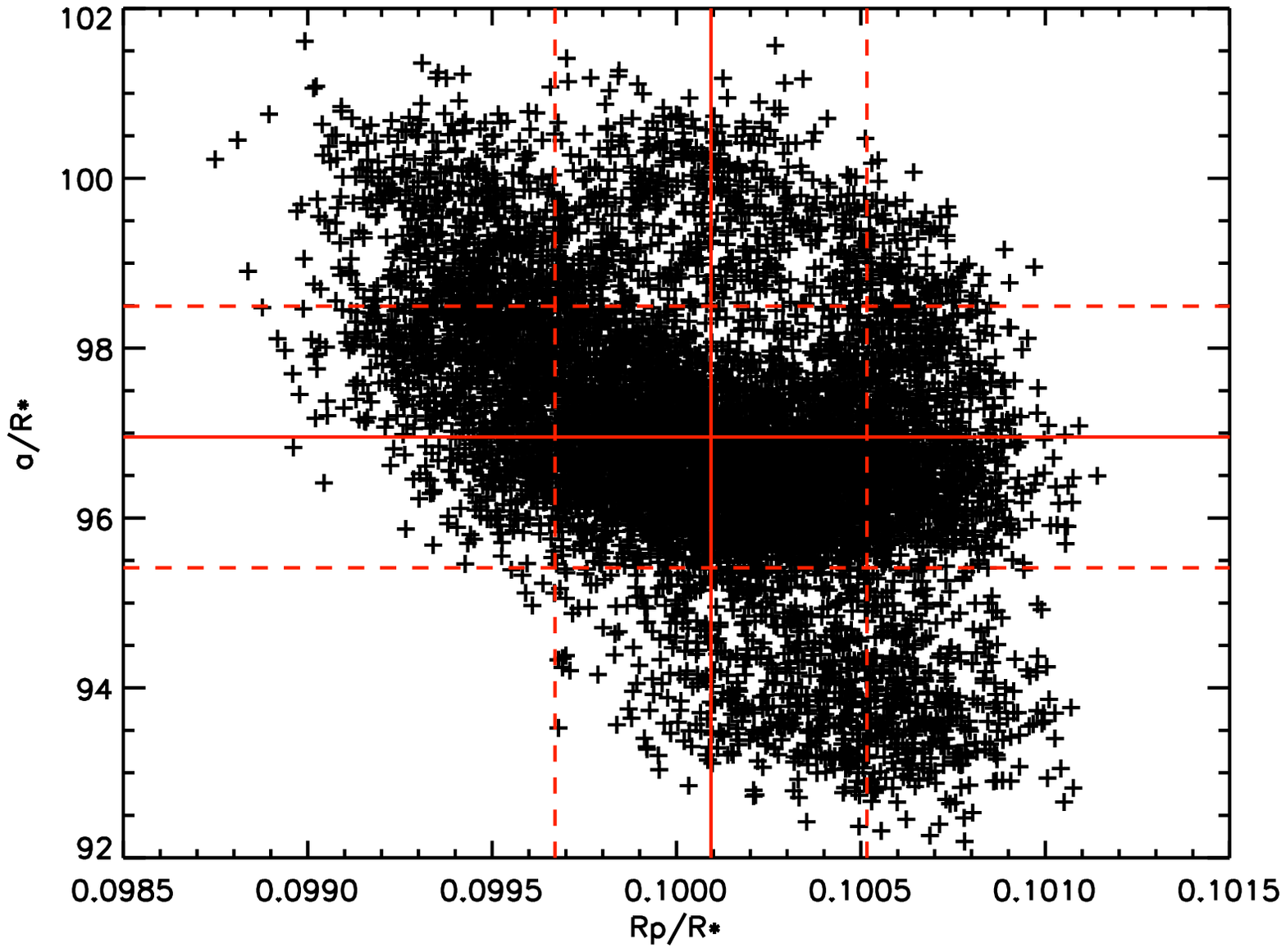}
\includegraphics[scale=0.52]{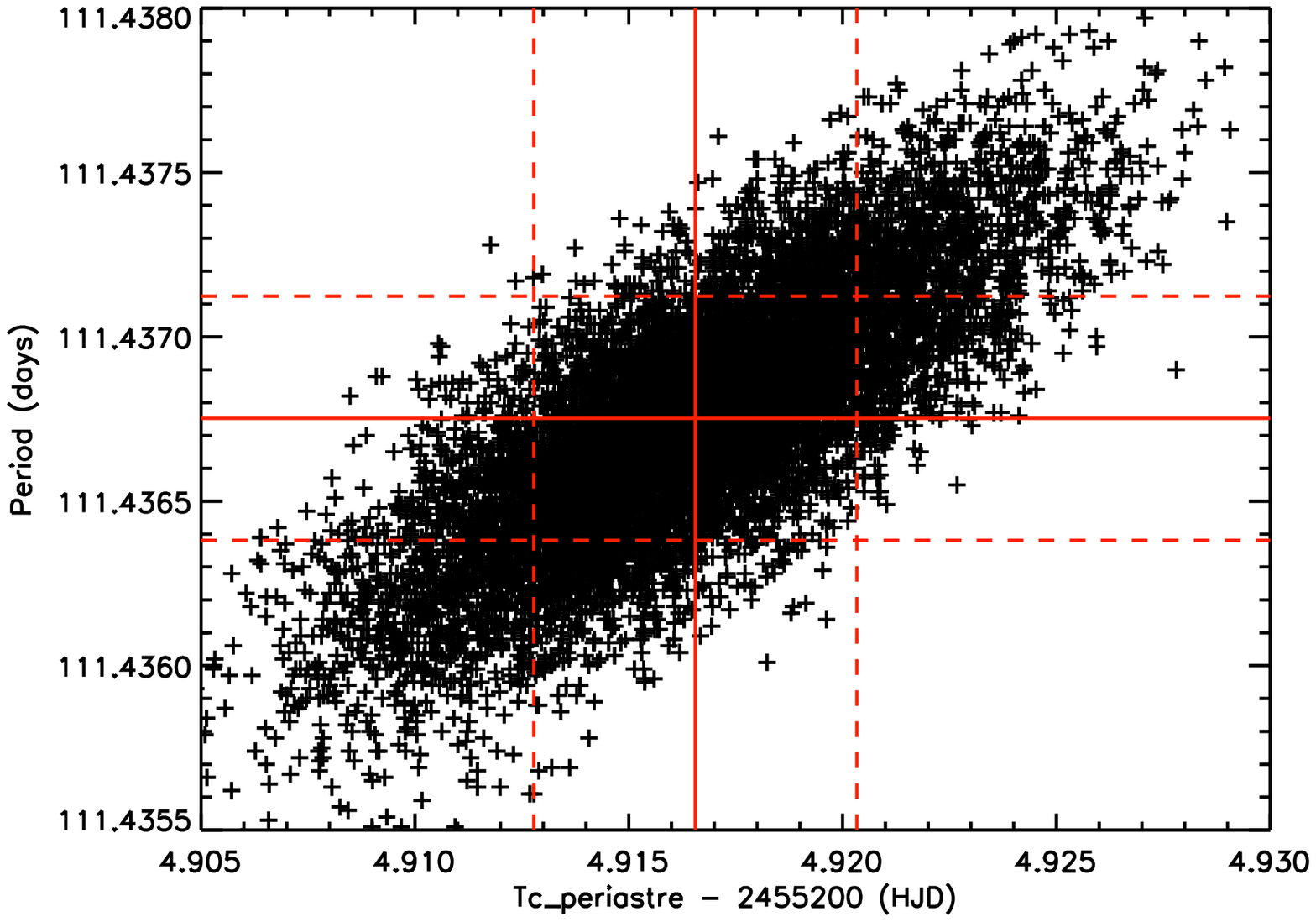}
\vspace{0.8cm}
\caption{Distributions of the parameters obtained for the 
$15000$ fits performed with the prayer bead and bootstrap (Sect.~\ref{sect_best_parameters}).
As example, four parameters are plotted here: $a/R_*$ vs. $R_\mathrm{p}/R_*$ and 
$P$ vs. $T_0$(periastron). The mean values and the standard deviations are~shown.}
\label{fig_dist_prayer}
\end{center}
\end{figure}

We used the Prayer Bead method (Moutou et al.~\cite{moutou04}, Gillon et al.~\cite{gillon07})
as applied by D\'esert et al.~(\cite{desert09}) to compute the mean values of the free 
parameters and their statistic and systematic uncertainties (see examples in Fig.~\ref{fig_dist_prayer}), 
together with the Levenberg-Marquardt algorithm to provide the best fit at each iteration
of our procedure.
This method was applied to the \spitzer\ photometry in order to account for possible 
correlated noise in the error budget. We simultaneously applied a bootstrap procedure to 
the radial velocity measurements, after having quadratically added a systematic uncertainty to  
the radial velocity data sets in order to put to unity their corresponding reduced $\chi^2$. 
Thus we quadratically  added 12~\ms\ 
to the \textit{ELODIE} uncertainties, and 5.0 and 1.7~\ms\ to the \textit{HIRES} uncertainties (pre- 
and post-upgrade, respectively). 
In total, $15000$ shifts and fits of transit light curves and bootstraps of radial velocity errors were
produced to derive the set of parameters and to extract their means and their corresponding 
standard deviations. In addition, we performed additional fits with the light curve
considering data starting at ten 
different epochs before the ingress; by performing a prayer bead on $1500$ shifts for each of 
those ten fits, we could estimate the errors caused by the uncertainty in the shape of the 
out-of-transit baseline (see Sect.~\ref{sect_spitzer_data_reduction}). 

The averaged values we obtained for the 
seven fitted parameters linked to the \spitzer\ light curve extraction 
(Sect.~\ref{sect_spitzer_data_reduction})
are:
$A_1 = 58818$ and 
$A_2 = 6.860$  
for the baseline, and  
$K_1 = 0.2142$, 
$K_2 = 0.1538$, 
$K_3 = 0.1083$, 
$K_4 = 0.0729$, and
$K_5 = 0.0050$
for the pixel-phase  effect. However, those 
parameters could be different at different epochs and for different pixel locations, so this 
is not clear whether the $A_j$ and~$K_i$~values we derived here could be applied to other IRAC observations with Warm-\spitzer.

The best solution of the combined fit is plotted in Fig.~\ref{fig_transit_all}, 
with a ramp cut up to HJD\,$=-2\,455\,210.325$, i.e. the first 1.8~hour of the 
observation was discarded, corresponding to the first 15000 over the 
159232 unbined frames.  The upper panel of Fig.~\ref{fig_transit_all} 
shows the radial velocities and the lower panel shows 
the \spitzer\ photometry binned per 250~pixels (the middle panel shows 
the Rossiter-McLaughlin fit that is discussed below in Sect.~\ref{sect_RM}).

The derived parameters are reported in Table~\ref{table_parameters} together with 
their error bars; they are ranked  as a function of 
the way they are derived. First are reported the free parameters of the combined fit that are 
listed in Sect.~\ref{sect_free_parameters} and which are constrained by photometry only, 
then those that are constrained both by radial velocities and photometry,
and then those  constrained by radial velocities only. In this last category, 
the dispersion around the obtained radial velocity shifts are also reported. 
Twelve adjusted parameters of the combined fit are reported here. 
Then Table~\ref{table_parameters} shows the parameters that are directly derived from the 
above free parameters of the fit, without any additional hypothesis. This includes the transit 
and eclipse timings, the latest being obtained from Laughlin et al.~(\cite{laughlin09})
and Gillon~(\cite{gillon09a}).
The following parameters are those that are derived by assuming the stellar mass 
$M_* = 1.01 \pm 0.05$~\MS\ (Sect.~\ref{sect_free_parameters}) together with the 
parameters derived above and through the Kepler third law. The last parameter set 
in Table~\ref{table_parameters} are those relative to the Rossiter-McLaughlin fit
that are obtained below in Sect.~\ref{sect_RM}.

\begin{figure}[h] 
\begin{center}
\includegraphics[scale=0.37,angle=90]{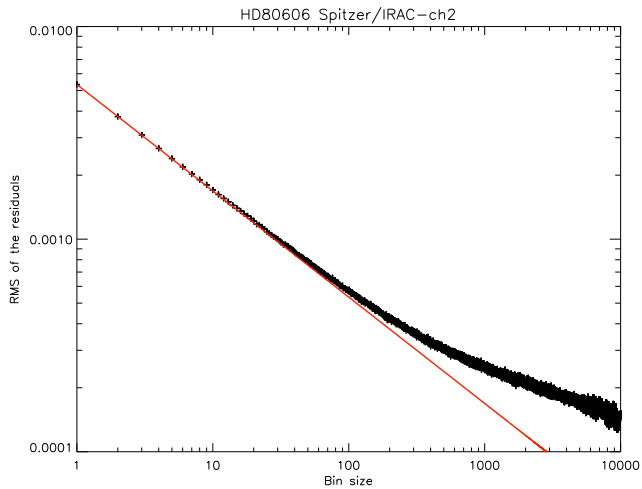}
\caption{Root mean square (RMS) of binned residuals of the \spitzer\ transit light curve as a 
function of the bin size $n$. The solid, red line is proportional to $n^{-1/2}$, as expected if 
only photon noise is considered.}
\label{fig_bin_residuals}
\end{center}
\end{figure}

The dispersion of the \spitzer\ photometry around the transit light curve fit represents 
$5.3 \times 10^{-3}$ of the stellar flux for unbinned frames.
This is the expected level of the photon noise.
The amplitude of the correlated noise, as seen for different bin-sizes of the light curve, 
is of the order of $1.4 \times 10^{-4}$ 
of the stellar flux (Fig.~\ref{fig_bin_residuals}).
A bump in the light curve with an amplitude of $\sim1$~mmag 
is seen just before the transit mid-time (see Sect.\ref{sect_warm_spitzer_LC}). 
It could be instrumental or due to a spot (see Sect.~\ref{sect_warm_spitzer_LC}).
We performed fits without taking into account these 
points; this did not change significantly the derived parameters.

\subsection{Rossiter-McLaughlin fit}
\label{sect_RM}

The radial velocity measurements secured during 
transits have been fitted in order to measure the sky-projected angle $\lambda$ between 
the planetary orbital axis and the stellar rotation axis. The data we use are 
the \textit{SOPHIE} observations of the February-2009 transit 
(Moutou et al.~\cite{moutou09}) and the new ones we secured in 
January 2010 (Sect.~\ref{section_sophie}). We also used  
the \textit{HIRES} data of the June-2009 transit (Winn et  al.~\cite{winn09a}).

We first measured the radial velocity shift of each data set by comparison to the 
center-of-mass radial velocity references computed in Sect.~\ref{sect_combined_fit}. 
Radial velocity measurements performed near but off the transit are mandatory 
to constrain those shifts. A lack of such reference observations could prohibit 
an accurate measurement of the spin-orbit angle.
For example in the case of the recent Rossiter-McLaughlin observation of 
the planetary system Kepler-8 
(Jenkins et al.~\cite{jenkins10}), the paucity of off-transit observations makes 
difficult to conclude if the apparent asymmetry of the Rossiter-McLaughlin
shape by comparison to the Keplerian curve is due to an actual spin-orbit 
misalignment or to a shift due to another cause, as stellar activity and/or instrumental~drifts.

\begin{figure}[h] 
\begin{center}
\includegraphics[scale=0.58]{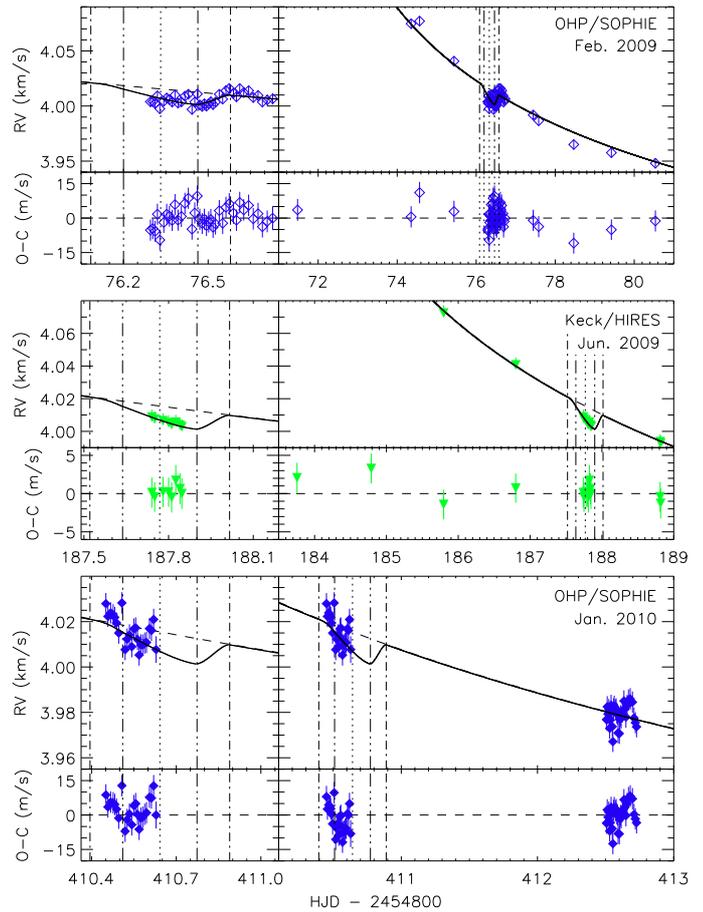}
\caption{Radial velocities of \cibleb\ around transits and their Ros\-siter-McLaughlin fit.
The upper panel shows the \sophie\ observations around the 14 February 2009 transit (Moutou 
et al.~\cite{moutou09}), the middle panel shows the \textit{HIRES} observations around the 
5 June 2009 transit (Winn et al.~\cite{winn09a}), and the lower panel shows the \sophie\ 
observations around the 13-14 January 2010 transit (Sect.~\ref{section_sophie}).
On each of these three panels the data are plotted together with the 1-$\sigma$ error 
bars. The fits with and without transit are the solid and dashed lines, respectively.
The left panels show the data during the nights of the transits, and the right panels 
show all the data secured the nights before and after the transits to allow the 
measurement of the radial velocity shift of each data set. The O-C residuals of the 
fit including the  Rossiter-McLaughlin anomaly are also plotted.
Moreover on all panels, the vertical dotted lines show the mid-transit, the vertical dot-dashed 
lines show the first and fourth contacts, and the vertical dot-dot-dashed lines show the second 
and third~contacts.}
\label{fig_RM}
\end{center}
\end{figure}

In the case of \cible,  we used as reference for the February-2009 data the nine \sophie\ 
measurements performed the nights before and after the transit night, as well 
as the nine ones performed after the fourth contact the night of the transit. 
For the 2010 \sophie\ data we used the 33 exposures obtained the January-15 night, 
i.e. two days after the transit.
Finally, for the \textit{HIRES} data we used the six June-2009 measurements secured before 
and after the night of the transit (the June-2009 data were excluded from the \textit{HIRES} 
post-upgrade data set used in Sect.~\ref{sect_combined_fit} for the fit of the orbit).
The radial velocity shift for the three data sets are reported in Table~\ref{table_parameters}.
We found a shift of $14.4 \pm 1.9$\,\ms\ between the 2009 and 2010 \sophie\ data sets, and 
a shift of $3.2 \pm 1.3$\,\ms\ between the  \textit{HIRES} measured in June 2009 and the 
other post-upgrade \textit{HIRES} data secured more than one year earlier.
These shifts could result from a combination of instrumental effects, unknown component 
in the system and/or activity-induced jitter.

The reference observations secured near each of the three  transits 
are plotted in the right panels of Fig.~\ref{fig_RM}. 
Using these off-transit radial velocities, we also computed the uncertainties we
had to add to the radial velocities tabulated error bars in order 
to put to unity the reduced $\chi^2$ corresponding to the Keplerian fit. 
We thus quadratically added 1.0\,\ms\ and 2.5\,\ms\ to the \sophie\ measurements 
secured in 2009 and 2010 respectively, and 1.7\,\ms\ to the \textit{HIRES}
measurements (in agreement with the other \textit{HIRES} post-upgrade 
data used above, see Sect.~\ref{sect_best_parameters}).

We model the Rossiter-McLaughlin anomaly shape using the analytical approach 
developed by Ohta et al.~(\cite{otha05}). The complete model has 14 parameters: 
the stellar limb-darkening linear coefficient $\epsilon$, 
the transit parameters $R_\mathrm{p}/R_*$, $a/R_*$ and $i$, 
the orbital parameters ($P$, $T_0$, $e$, $\omega$, $K$), 
the three radial velocity shifts measured above, 
and finally $V \sin I_*$ and $\lambda$.
We computed $\epsilon=0.722$ in the wavelength range $5300-6300$\,\AA\ 
using the same method as in Sect.~\ref{sect_ecc_LC}.
The transit and orbital parameters are fixed from the results obtained 
in Sect.~\ref{sect_combined_fit}; their uncertainties are negligible by 
comparison to those of the 
two main free parameters of~the~Rossiter-McLaughlin fit: $\lambda$, which 
is constrained by the asymmetry of the anomaly, and $V \sin I_*$, which is 
constrained by its~amplitude. 

\begin{figure}[h] 
\begin{center}
\includegraphics[scale=0.53]{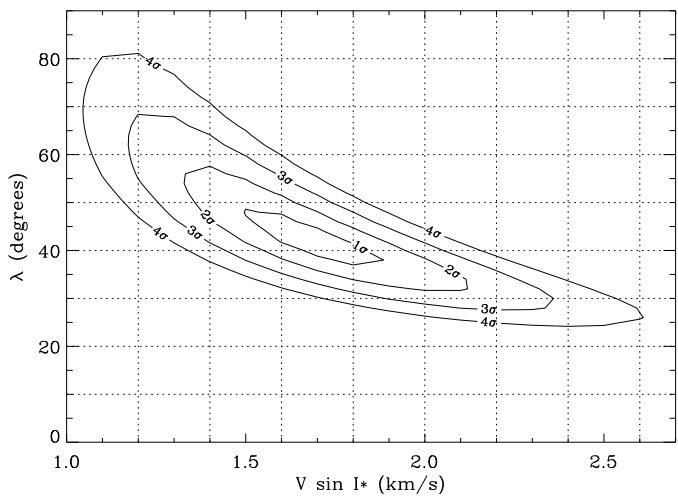}
\caption{Confidence interval contours for the $\lambda$ and $V \sin I_*$ values
from the Rossiter-McLaughlin fit.}
\label{fig_chi2_RM}
\end{center}
\end{figure}

The best fit is plotted in the middle panel of Fig.~\ref{fig_transit_all} and in 
Fig.~\ref{fig_RM}. It is obtained for $\lambda = 42^{\circ}$ and 
$V \sin I_* = 1.7$\,\kms; its $\chi^2$ is 117.8 for 117 degrees of freedom.
The confidence interval contours estimated from $\chi^2$ variations
(H\'ebrard et al.~\cite{hebrard02}) for the two correlated $\lambda$ and $V \sin I_*$
parameters are plotted in Fig.~\ref{fig_chi2_RM}. The uncertainties obtained this 
way are $\pm6^{\circ}$ and $\pm0.2$\,\kms, respectively. We increased them in 
order to take into account for the uncertainties in the radial velocity shifts of the 
three data sets with respect to the Keplerian curve computed in Sect.~\ref{sect_combined_fit}.
Our final values are $\lambda = 42^{\circ} \pm 8^{\circ}$ and $V \sin I_* = 1.7 \pm 0.3$\,\kms.
We checked that 
the uncertainties on the parameters derived in Sect.~\ref{sect_combined_fit} and used 
here in the Rossiter-McLaughlin fit imply negligible uncertainties on $\lambda$
and $V \sin I_*$; 
similarly, the uncertainty in the $\epsilon$ limb-darkening coefficient is negligible.

The adopted values are reported in Table~\ref{table_parameters}, together with the 
dispersion of the radial velocities performed during the transit with respect to the 
Rossiter-McLaughlin fit. The dispersion is in the range $4-5$\,\ms\ for \sophie\ data; 
we have a similar dispersion around the Keplerian curve for the \sophie\ reference 
measurements performed before and after the transit. For the \textit{HIRES} data, 
the dispersion around the Rossiter-McLaughlin fit is $0.8$\,\ms, whereas it is two 
times larger ($1.7$\,\ms) for the June-2009 \textit{HIRES} measurements performed 
before and after the transit. So it could remain a slight systematic in the \textit{HIRES}
radial velocities secured during the transit. However this effect is small, and is 
detected here due to the particularly high accuracy of these measurements.

\section{Discussion}
\label{sect_discussion}

\subsection{Warm-\spitzer\ transit light curve}
\label{sect_warm_spitzer_LC}

This planetary transit observations is among the first ones secured with \spitzer\ 
in its post-cryogenic mission. This shows that accurate transit light curves
can be obtained in this second part of~the observatory mission 
despite enhanced ramp and pixel-phase~effects.
Our Warm-\spitzer\ transit light curve of \cible\ has 
lower uncertainties than those obtained from 
the ground. The typical durations of the pixel-phase effect (70~minutes) and 
the planetary transit (12~hours) are on time-scales different enough 
 to avoid significant 
uncertainties on the derived system parameters due to this instrumental effect.
The transit light curve is well detected, without extra signatures of transiting 
rings or satellites. According to our accuracy, 
the signature of an hypothetic satellite would have been 
detected up to a magnitude depth slightly below 1~mmag; this corresponds 
to an upper limit of 2\,R$_\oplus$ on the~radius.

A bump in the light curve is seen just before the transit mid-time with an 
amplitude of $\sim1$\,mmag. It could corresponds to the planet occulting 
a dark spot during the transit (Pont et al.~\cite{pont07}, Rabus et al.~\cite{rabus09}).
\cible\ is not an active star, so the probability of such event is low.
The stability at the level of a few mmag of the stellar flux observed in the optical 
(Sect.~\ref{sect_ground_photom}) would correspond to a 
stability below 1~mmag at 4.5\,$\mu$m, assuming dark spots are 1000-K cooler than 
the stellar surface (D\'esert et al.~\cite{desert10}); so the optical 
photometry cannot exclude 
that the feature detected in the \spitzer\ light curve is due to a spot.
One can note that at the time of this feature (near 137~hours after the 
periastron in Fig.~\ref{fig_transit_all}), a simultaneous feature is also 
seen in the  \sophie\ radial velocity data. This could argue 
in favor of the interpretation of this event in term of phenomenon at 
the surface of \cible. This seems however unlikely, since a photometric 
feature with this low flux amplitude should a priori not produce such high radial 
velocity effect. The radial velocity feature is more likely an instrumental 
systematic. 
The feature in the \spitzer\ light curve is detected at a maximal value 
of the pixel-phase effect (Fig.~\ref{fig_spi}) so we should remain cautious
about instrumental effects. We monitored the coordinates of the 
target on the subarray through the observation in order to test if the bump 
corresponds to a particular area of the detector. This is not the case: 
at the epoch of the bump, the target is located on a position of the 
detector where the target passes numerous times before and after this 
event, and where the pixel-phase effect apparently is well corrected 
by our procedure. So we found no strong reasons to particularly 
favor an instrumental effect to explain the presence of this feature 
in the \spitzer\ light curve. The most likely explanation is the presence 
of a small star spot on the surface \cible, above which the planet is transiting.

Our data set results from the first joined campaign during which 
space-based photometry and high-precision radial velocities were 
carried out simultaneously. If large stellar spots are detected during 
a similar campaign, this would help to better understand the effect of 
stellar activity on radial velocity measurements. Stellar jitter indeed 
introduces noise in measured radial velocities and is a significant 
limitation in high-precision velocimetry (see, e.g., 
Saar \&\ Donahue~\cite{saar97}, Santos et al.~\cite{santos00}, 
Boisse et al.~\cite{boisse09}, Queloz et al.~\cite{queloz09}, 
H\'ebrard et al.~\cite{hebrard10a}). No stellar spots
large enough to be spectroscopically detected 
were apparently seen~here.

\subsection{Comparison with previous measurements}
\label{sect_compariton}

The \cible's system parameters that we report in Table~\ref{table_parameters}
have better accuracy by comparison with previous studies 
(Moutou et al.~\cite{moutou09}, Pont et al.~\cite{pont09}, Gillon~\cite{gillon09a}, 
Winn et al.~\cite{winn09a}, Hidas et al.~\cite{hidas10}). With respect to 
the ground observation of nearly all the transit phases used by Winn et al.~(\cite{winn09a}), 
the uncertainties presented here are better by factors two to five. 
Exceptions are the parameters $e$, $\omega$ and $K$ for which the error bars 
are not significantly reduced. This is due to the fact that most of the 
constraints on these three parameters come from the radial velocity 
on a 9.5-year time span and already used by previous~studies.

For most of the parameters, the revised, more accurate values are in agreement 
within $1\,\sigma$ with those previously published. 
There is a $2-\sigma$ disagreement on the inclination $i$ 
of the orbit, which is found in our study to be slightly lower; this 
implies a more grazing transit, a larger impact parameter $b$, and a 
slightly longer duration for ingress and egress ($\sim10$~minutes longer).
The $a/R_*$-ratio is also found to be 5\,\%\ smaller, a shift by $2\,\sigma$  according 
the error bars from Winn et al.~(\cite{winn09a}).

The projected stellar rotational velocity we found, $V \sin I_* = 1.7 \pm 0.3$\,\kms,
is slightly larger than the value obtained by Winn et al.~(\cite{winn09a}) from the 
Rossiter-McLaughlin fit of their \textit{HIRES} data: $V \sin I_* = 1.12^{+0.44}_{-0.22}$\,\kms.
Whereas we also used their \textit{HIRES} in our analysis, our different result 
is explained by the fact that we use additional \sophie\ data and we measured a
smaller $R_\mathrm{p}/R_*$ ratio (see Sect.~\ref{sect_radius}). In addition, 
a difference with the study by Winn et al.~(\cite{winn09a}) is the shift of $3.2 \pm 1.3$\,\ms\ 
that we found between the \textit{HIRES} data secured near the transit and the other 
ones (Sect.~\ref{sect_RM}); this implies a slightly larger amplitude for the 
Rossiter-McLaughlin anomaly, thus a larger $V \sin I_*$.
The value $V \sin I_* = 1.7 \pm 0.3$\,\kms\ that we measured nonetheless 
agrees with those found from synthetic spectral fitting: $V \sin I_* = 1.8 \pm 0.5$ 
and $2.0 \pm 0.5$\,\kms\ by Valenti \& Fischer~(\cite{valenti05}) and Winn 
et al.~(\cite{winn09a}), respectively. 
The discrepancy noted and discussed by several authors (Winn et al.~\cite{winn05}, 
Triaud et al.~\cite{triaud09}, Hirano et al.~\cite{hirano10}, Simpson et al.~\cite{simpson10})
between the $V \sin I_*$ measured from the Rossiter-McLaughlin effect and from the
spectral modeling of line broadening is negligible for such slow-rotating star like~\cible.

The stellar radius we measured, $R_\star = 1.007 \pm 0.024$\,R$_\odot$, is slightly 
larger than this obtained by Winn et al.~(\cite{winn09a}), namely 
$R_\star = 0.968 \pm 0.028$\,R$_\odot$. Both values remain in agreement 
with the one deduced by Moutou et al.~(\cite{moutou09}) from relationships between stellar 
radius, luminosity, temperature, gravity and mass, namely $R_* = 0.98 \pm 0.07$\,R$_\odot$. 

The radius ratio $R_\mathrm{p}/R_*$ and the timing of the transit that we found 
significantly differ from the values derived from ground-based observation;
this is discussed below.

\subsection{Planetary radius}
\label{sect_radius}

The radius ratio we found from the transit observed with \spitzer\ is 
$R_\mathrm{p}/R_* = 0.1001 \pm 0.0006$. By comparison, Winn et al.~(\cite{winn09a})
measured $R_\mathrm{p}/R_* = 0.1033 \pm 0.0011$; Pont et al.~(\cite{pont09})
found the same  radius ratio than Winn et al.~(\cite{winn09a}) but with an 
uncertainty three time larger, mainly due to the lack of ingress observation.
The radius ratio measured in the infrared from space is thus $\sim3$\,\%\ smaller 
than the one measured in the optical from ground. This is a  $3-\sigma$ difference 
according the error bar reported by Winn et al.~(\cite{winn09a}). The two horizontal 
dotted lines in the lower panel of Fig.~\ref{fig_transit_all} show the absorption depths 
expected with the two values; the infrared radius ratio clearly is smaller than the optical one.

The uncertainties in the background measurement (Sect.~\ref{sect_spitzer_data_reduction}) 
are too small to account for such a radius difference.
A variation of the stellar brightness due to spots could explain time-variations 
in the measured radius ratio (D\'esert et al.~\cite{desert10}). The upper limit of the 
brightness variations of the non-active star \cible\ (Sect.~\ref{sect_ground_photom})
is however too small to explain a $\sim3$\,\%\ radius-ratio variation.
The planetary thermal emission could produce an underestimation of 
the measured radius ratio (Kipping \& Tinetti~\cite{kipping09}). The emission 
level measured by Laughlin et al.~(\cite{laughlin09}) at 8\,$\mu$m near the
periastron is larger than the planetary emission expected at 4.5\,$\mu$m
at the transit; even with this overestimated planetary flux, the effect would be negligible 
according the error bars on the radius ratio we measure at 4.5\,$\mu$m.

The question of possible 
interpretation in terms of differential atmospheric absorption could be raised. 
The planet apparent radius as a function of wavelength has been shown 
to follow the equation $dR_p/d\lambda=H d\ln \sigma/d\lambda$ 
(Lecavelier des Etangs et al.~\cite{lecavelier08a},~\cite{lecavelier08b}), 
where $\lambda$ 
is the wavelength, $\sigma$ is the cross section of the mean 
atmospheric absorber, $H=kT/\mu g$ is the atmospheric scale height, 
$k$ is the Boltzmann constant, $T$ is the temperature, 
$\mu$ is the molecular mass, and $g$ is the planet gravity 
which is about 100\,m\,s$^{-2}$. If the variation of radius is due 
to variation of absorption by haze in the atmosphere as in the case 
of the HD\,189733b (Pont et al.~\cite{pont08}), even assuming 
Rayleigh scattering which produces the steepest variation of absorption 
as a function of wavelength, a temperature of about 5\,000\,K is 
required to interpret the present measurements, 
much higher than actually measured (Laughlin et al.~\cite{laughlin09}).
Indeed, because of the large planetary mass, at a typical temperature 
of 1000\,K the scale height is only 60\,km. The difference of measured 
radius is about 40~times this scale height, and is therefore unlikely 
due to atmospheric differential absorption. 

We found no astrophysical interpretations able to explain the radius difference  
between optical and infrared wavelengths. It is more likely that the error bars 
are slightly underestimated, maybe in the ground-based composite light curve.
We note that as we obtain a larger stellar radius than Winn et al.~(\cite{winn09a}), 
we obtain a similar planetary radius despite the different radius ratio: 
$R_\textrm{p} = 0.981 \pm 0.023$\,R$_\mathrm{Jup}$.

\subsection{Transit timing}
\label{sect_ttv} 

The mid-times of the transit and the eclipse given in Table~\ref{table_parameters} 
are those that are \textit{measured}. This means that the transit mid-time $T_t$
we report is in significant advance by about 2.5~minutes in comparison to this predicted 
from the epoch $T_0$ we report for the periastron of the planet 
(see Sect.~\ref{sect_ecc_LC}). Indeed, this latest time is in the referential frame of 
the radial velocities, which are these of the star.
Similarly, the epoch $T_e$  of the eclipse reported in Table~\ref{table_parameters} 
is delayed by about 15~seconds by comparison to the ephemeris computed 
from~$T_0$

The mid-time of the transit we obtain is accurate at the level of 1.5~minute, 
whereas the accuracy is 6~minutes for the epoch of the periastron. Laughlin 
et al.~(\cite{laughlin09}) obtained a 4-minute accuracy for the mid-time 
of the eclipse. These accuracies are high, especially when they are compared with 
the long duration of the transit and the long orbital period. 
Using radial velocities alone, even if they span a time as long as 9.5 years, 
the error bars on the eclipse and periastron timing are three times larger.
For the transit timing, the radial velocities can not predict it at better than 
two to three hours. Transit and eclipse detections allow here more accurate timings.

This high-accuracy offers opportunity to look for possible transit timing 
variations (TTVs). According to the mid-transit time and the period we found, 
we obtain for the February and June~2009 transits a mid-time that is $\sim23$\,minutes 
earlier than the $T_t$ time measured by Winn et al.~(\cite{winn09a}) from these events.
Their accuracy on this timing was $\pm7$\,minutes, 
so the disagreement is at the level of $3\,\sigma$.
If it is not caused by underestimated systematic uncertainties, 
such a difference in the transit timing could in principle be due to the presence 
of additional bodies in the system, such as satellites of the transiting planet or 
additional planetary-mass bodies in the system 
(see, e.g., Holman \& Murray~\cite{holman05},
Agol et al.~\cite{agol05}, Nesvorny \& Beaug\'e~\cite{nesvorny10}).

We explored a small region of the parameter space of a hypothetical additional 
planet in search for a few examples that could explain a $\sim20$-minute difference 
between two nearby transits. For this,  we performed a series of 3-body 
simulations by integrating the equations of motion using the Burlisch-Stoer 
algorithm implemented in the Mercury6 package (Chambers~\cite{chambers99}). 

Planets in resonant orbits could explain such TTV even with masses low 
enough to prohibit their detection with the available radial velocities. 
For example, a 15-Earth-mass
planet in a circular 4:1-resonant orbit 
produces the adequate timing anomalies, with many pairs of transits 
exhibiting a $\sim20$-minute discrepancy. The simulated timing variations 
exhibit an amplitude of 80~minutes in this case. A 0.17-Jupiter-mass planet 
at the 6:1 resonance produces similar results, although the overall amplitude 
is reduced to $\sim40$~minutes. Both cases would imply radial velocity 
variations with semi-amplitudes $K\simeq1.5$ and 4\,\ms, respectively, 
so we cannot exclude those two cases: the first one would be undetectable 
in the available radial velocity data set, whereas the second one would be at 
the limit of detection. 
More massive planets on non-resonant orbits could be excluded. For example, 
a planet located in a circular orbit at 3~AU and having a mass of 1.7~Jupiter 
masses would imply a small number of pairs of nearby transits exhibit a 
variation of the order of 20~minutes. It would imply radial velocity 
variations of $K\simeq30$\,\ms\ which is detectable with the available 
radial velocities.
In all these cases, the short-term 
variations (in timescales of the order of 2000~days) exhibited by the orbital 
parameters of the transiting planet remain below the level of precision 
reached from our combined fit (Sect.~\ref{sect_combined_fit}).
However, the variations in timescales of the order of $10^{5-6}$~days are substantially 
larger in all cases. In particular, some 4:1-resonant cases could be unstable 
after a few thousands of years. 

We also explored the possibility that the observed discrepancy were 
produced by a satellite to the transiting planet. In the more favorable 
case of having a satellite orbiting the planet at the Hill's radius, we found 
that its mass must be around 3/100 that of \cibleb\  
for it to be the cause of a $\sim20$-minute delay in the transit occurrence, 
corresponding to 40~Earth masses.

\subsection{Spin-orbit misalignment}
\label{sect_discuss_RM}

We confirm the spin-orbit misalignment in the \cible\ system and reduce 
the uncertainty in its measurement: $\lambda=42\pm8^{\circ}$. This allows 
scenarios with aligned spin-orbit to be rejected with a high-level of confidence.
Such misalignment would imply an asymmetry in the Rossiter-McLaughlin 
anomaly as the star is not a perfect sphere but is slightly elongated at the 
equator due to its rotation. However, the accuracy of the data is not high enough to 
allow such tiny effect to be detected.

The $\lambda$ angle measures the sky-projected angle between the planetary orbital 
axis and the stellar rotation axis. Its actual value remains unknown, as the 
inclination $I_*$ of the stellar rotation axis is undetermined. In cases where 
$\lambda = 0$ is measured, it  is reasonable to assume $I_* \simeq 90^{\circ}$
and an actual spin-orbit alignment. In cases as \cible\ where $\lambda$ is 
significantly different from 0, there is certainly no reasons to assume the 
stellar rotation axis is parallel to the sky plan.

The projected stellar rotational velocity we get from the Rossiter-McLaughlin 
fit is $V \sin I_* = 1.7 \pm 0.3$\,\kms. According to our measured 
radius for \cible, $R_\star = 1.007 \pm 0.024$\,R$_\odot$, and to the relation 
from Bouchy et al.~(\cite{bouchy05}), this $V \sin I_*$ translates into a stellar rotation 
period $P_\mathrm{rot}/\sin I_* = 30 \pm 5$\,days. So the stellar rotation 
period apparently is shorter than 40\,days. This is shorter than 
the limit $P_\mathrm{rot} > 50$~days we get in Sect.~\ref{sect_sophie_reduction}
from the low activity of \cible. This estimation based on the activity, however, is not 
accurate especially for long rotation periods. These two values suggest 
a rotation period in the range $40-50$~days, so a $I_*$-angle near $90^{\circ}$. 
The value $\lambda=42\pm8^{\circ}$ that we measure is thus 
probably close to the value of the 
misalignment angle without projection effect. 
So the actual angle between the spin-axis of \cible\ and the normal to the planetary 
orbital plane is about~$40^{\circ}$.

\begin{figure}[h] 
\begin{center}
\includegraphics[scale=0.52]{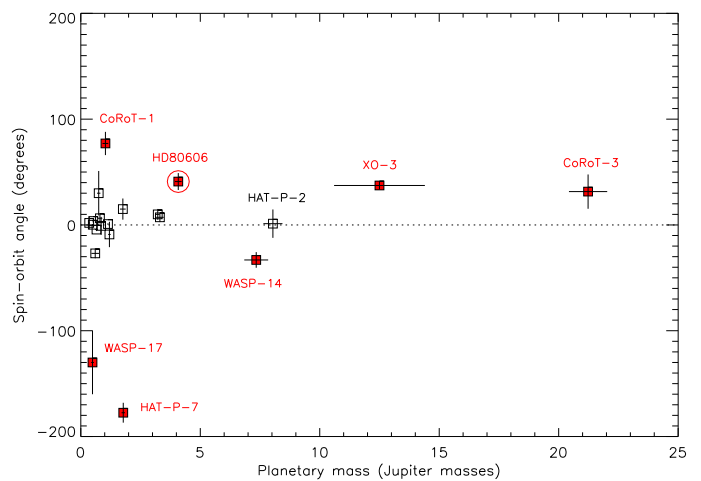}
\caption{Sky-projected $\lambda$-angle between the planetary orbital axis and the stellar rotation axis
as a function of the planetary mass, for 21 
published systems (see references in Sect.~\ref{sect_discuss_RM}).
Systems with $|\lambda | > 30^{\circ}$ are in filled, red symbols. \cible\ is marked by a circle.
}
\label{fig_revue_RM}
\end{center}
\end{figure}

The first case of a planetary system with a stellar spin misaligned 
with the normal of the planetary orbit was discovered by H\'ebrard et 
al.~(\cite{hebrard08}) 
in the XO-3 system. This result was hereafter confirmed by Winn 
et al.~(\cite{winn09b}), who however found a lower $\lambda$-value.
\cible\ was the second system reported to 
have a spin-orbit misalignement by Moutou et al.~(\cite{moutou09}).
This result was subsequently confirmed and refined successively by 
Pont et al.~(\cite{pont09}), Gillon~(\cite{gillon09a}), Winn et al.~(\cite{winn09a})
and finally by the present study.
Other planetary systems with significantly misaligned spin-orbit have been 
reported since then: 
WASP-14 (Johnson et al.~\cite{johnson09}), 
WASP-17 (Anderson et al.~\cite{anderson10}), 
HAT-P-7 (Winn et al.~\cite{winn09c}, Narita et al.~\cite{narita09a}), and
CoRoT-1 (Pont et al.~\cite{pont10}). 
There are thus now six known misaligned systems.
Three other systems may be misaligned, but the large uncertainties in the 
reported $\lambda$-values prohibit definitive conclusion: 
WASP-3 (Simpson et al.~\cite{simpson10}), 
TrES-1 (Narita et al.~\cite{narita07}), and 
CoRoT-3 (Triaud et al.~\cite{triaud09}).
The recent case of Kepler-8 (Jenkins et al.~\cite{jenkins10}) is presented
as moderately misaligned but this requires confirmation (see Sect.~\ref{sect_RM}).
On the other hand, eleven systems are apparently aligned, namely
HD\,209458 (Queloz et al.~\cite{queloz00}, Wittenmyer et al.~\cite{wittenmyer05}), 
HD\,189733 (Winn et al.~\cite{winn06}, Triaud et al.~\cite{triaud09}, Collier Cameron et al.~\cite{cameron09}), 
HD\,149026 (Wolf et al.~\cite{wolf07}, Winn \& Johnson in prep.),
HD\,17156 (Cochran et al.~\cite{cochran08}, Barbieri et al.~\cite{barbieri09}, Narita et al.~\cite{narita09b}),
HAT-P-1 (Johnson~\cite{johnson08}), 
HAT-P-2 (Winn et al.~\cite{winn07}, Loeillet et al.~\cite{loeillet08}), 
HAT-P-13 (Winn et al.~\cite{winn10b}),
WASP-6 (Gillon et al.~\cite{gillon09b}), 
TrES-2 (Winn et al.~\cite{winn08}), 
TrES-4 (Narita et al.~\cite{narita10}), and
CoRoT-2 (Bouchy et al.~\cite{bouchy08}).

Fabrycky \& Winn~(\cite{fabrycky09}) have shown that the whole sample of the 
measured $\lambda$-angles could be well reproduced with a bimodal distribution, 
by assuming that a fraction of the orbits have random orientations relative to the stars 
and the remaining ones are perfectly aligned. The systems of misaligned orbits 
would be those which experienced gravitational interactions between planets 
and/or stars, such as Kozai migration. 

In Fig.~\ref{fig_revue_RM} we show the measured $\lambda$-angles as a function 
of the planetary masses for the 21 systems with published measurements reported above.
This plot suggests a scenario with three distinct populations.
Indeed, for the planets with masses similar to that of Jupiter, most of the spins are 
aligned with the orbits; this is expected for planets that formed in a protoplanetary disk far 
from the star and that slowly migrated closer-in at a later time.
A small fraction of these Jupiter-mass
planets however exhibits large $\lambda$-angles; this could be the signature of the 
second population in the scenario by Fabrycky \& Winn~(\cite{fabrycky09}), these 
which experienced gravitational interactions and that apparently are less frequent.
These planets also seem to be those than can have extreme $\lambda$-values.

A third population could be formed by the large-mass planets. Indeed, most of them 
exhibit a misaligned spin-orbit (see also Johnson et al.~\cite{johnson09}), 
maybe all of them\footnote{The only exception, 
HATP-P-2, could actually be also misaligned as the inclination of the 
stellar rotation axis is unknown.}.
This is a priori surprising as one can expect the most 
massive planets are the ones for which exciting the inclination is more difficult.
If indeed such misalignments are frequent for the high-mass-planet
population, this suggests a different evolution scenario for them. 
Maybe the more massive planets could not really slowly migrate because 
of the interactions with the disk. In that case, only more severe interactions with 
another planet or a star could be the cause of migration for massive planets, such 
interactions also affecting the inclination of the orbit.
We also note that the $\lambda$-angles seem lower for the massive planets 
than those of the misaligned planets with lower masses; this suggests as 
well a different scenario for the processes able to modify the inclination of low- 
and large-mass~planets. 

Rossiter-McLaughlin observations of other systems with transiting 
massive planets should be performed to confirm or not that there 
are preferentially tilted.

\section{Conclusion}
\label{sect_conclusion}

We presented an observation of the 12-hour-long transit of the highly eccentric, 
111.4-day-period exoplanet \cibleb\ performed in January~2010.
The~transit light curve we present is among the first ones carried with the 
post-cryogenic \spitzer. Its shows systematic effects stronger than those seen 
in the Cold-\spitzer\ but the accuracy remains clearly better that ground-based observations.
Together with the \sophie\ measurements acquired at the same epoch, this is 
one of the first  observational campaigns performed simultaneously in radial 
velocities and space-based high-accuracy photometry. 
With previously available data sets, this allows the parameters of this system 
to be significantly refined thanks to combined fits. 
There is a possible detection of a variation in the transit timing, which has  
to be confirmed by additional observations of \cibleb\ transits, from ground or space.
A dark spot was also possibly detected on the surface of this inactive star.
The spin-orbit misalignment is clearly confirmed and the $\lambda$ angle is accurately 
measured. As most of the massive planets for which this angle is measured, the orbit 
of \cibleb\ is misaligned with the equatorial plan of its host star.
This suggests a separate evolution scenario for massive planets in comparison with 
Jupiter-mass planets.

\begin{acknowledgements}
We thank the Haute-Provence Observatory 
staff that allowed \sophie\ observations despite the ongoing 
break of the 193-cm telescope, and in particular Fabien Fillion, Fran\c{c}ois Moreau, 
Jacky Taupenas, Jean-Pierre Troncin and St\'ephane Favard.
We are grateful to C.~Adami, D.~Russeil and M.~Dennefeld for their flexibility in 
the OHP 120-cm photometry observations, M.~Gillon and J.~N.~Winn 
for useful discussions, and the referee G.~Laughlin.
We thank the \spitzer\ Director for having awarded us Discretionary Time, and
the \spitzer\ staff and in particular Nancy Silbermann for their assistance in the 
preparation of the observation.
We acknowledge supports of the ``Programme National de Plan\'etologie'' (PNP) of CNRS/INSU, 
the Swiss National Science Foundation, 
and the French National Research Agency (ANR-08-JCJC-0102-01 and ANR-NT05-4-44463).
D.E. is supported by CNES. 
A.E. is supported by a fellowship for advanced researchers from the Swiss National Science Foundation.
N.C.S. would like to thank the support by the European Research Council/European 
Community under the FP7 through a Starting Grant, as well from Funda\c{c}\~ao para a 
Ci\^encia e a Tecnologia (FCT), Portugal, through a Ci\^encia\,2007 contract funded by 
FCT/MCTES (Portugal) and POPH/FSE (EC), and in the form of grants reference 
PTDC/CTE-AST/66643/2006 and PTDC/CTE-AST/098528/2008.

\end{acknowledgements}

\end{document}